# A Local Counter-Regulatory Motif Modulates the Global Phase of Hormonal Oscillations

Dong-Ho Park[1†], Taegeun Song[1†], Danh-Tai Hoang[1,2,3†], Jin Xu[1,4], and Junghyo Jo[1,4*]

[1]Asia Pacific Center for Theoretical Physics, Pohang, Gyeongbuk 37673, Korea

[2]Laboratory of Biological Modeling, National Institute of Diabetes and Digestive and Kidney Diseases, National Institutes of Health, Bethesda, Maryland 20892, United States of America

[3]Department of Natural Sciences, Quang Binh University, Dong Hoi, Quang Binh 510000, Vietnam

[4]Department of Physics, Pohang University of Science and Technology, Pohang, Gyeongbuk 37673, Korea

*Correspondence: jojunghyo@apctp.org

†These authors contributed equally to this work.

**Counter-regulatory elements maintain dynamic equilibrium ubiquitously in living systems. The most prominent example, which is critical to mammalian survival, is that of pancreatic α and β cells producing glucagon and insulin for glucose homeostasis. These cells are not found in a single gland but are dispersed in multiple micro-organs known as the islets of Langerhans. Within an islet, these two reciprocal cell types interact with each other and with an additional cell type: the δ cell. By testing all possible motifs governing the interactions of these three cell types, we found that a unique set of positive/negative intra-islet interactions between different islet cell types functions not only to reduce the superficially wasteful zero-sum action of glucagon and insulin but also to enhance/suppress the synchronization of hormone secretions between islets under high/normal glucose conditions. This anti-symmetric interaction motif confers effective controllability for network (de)synchronization.**

## Introduction

Living systems must maintain internal homeostasis in the face of external perturbations [1]. The endocrine system orchestrates dynamic equilibrium via long-range messengers known as hormones. Most physiological processes are controlled by negative feedback and antagonistic pairs of hormones, such as insulin/glucagon for glucose homeostasis [2], calcitonin/parathyrin for calcium homeostasis [3], and leptin/ghrelin for energy homeostasis [4]. Hormones also exhibit temporal oscillations [5,6]. The mechanistic reason for wave-like hormonal information broadcasting is less well known, although both a wave's amplitude and phase encode information. In addition to amplitude modulation, the relative phase coordination between different hormones and phase synchronization between different sources (cells or tissues) of the same hormones could have functional implications. Here, we focus on the example of phase modulation in glucose homeostasis, a much-studied case due to the critical importance of glucose homeostasis in human survival.

Glucose, which is a primary energy source in the body, is mainly regulated by glucagon and insulin secreted by α and β cells in the pancreas. Glucagon increases blood glucose by stimulating the breakdown of glycogen into glucose in the liver, while insulin decreases blood glucose by stimulating the synthesis of glycogen from glucose in the liver and clearing glucose into peripheral tissues. Pancreatic α and β cells are clustered together with an additional cell type, the δ cell, to form micro-organs known as the islets of Langerhans. Approximately one million islets are scattered throughout the human pancreas. Pancreatic α, β, and δ cells generate pulses of glucagon, insulin, and somatostatin, respectively. For the effective regulation of glucose homeostasis, coordination between insulin and glucagon pulses and between different islets is required. Approximate out-of-phase coordination of insulin and glucagon pulses has been observed in *in vitro* [7] and *in vivo* experiments [8]. This coordination implies interactions between islet cells. Indeed, it has long been observed that paracrine factors from α cells provide positive feedback to hormone secretion by β and δ cells, while δ cells provide negative feedback to α and β cells [9]. Furthermore, β cells exert negative feedback on α cells, but positive feedback on δ cells, which was recently confirmed [10] (See Fig. 1A). The rationale for this set of interactions is a long-standing puzzle [11]. A separate and equally important form of coordination is inter-islet synchronization. The independence or coherence of hormone secretions from one million islets must have a large impact on human physiology. It has been implicitly assumed that islets are always synchronized to produce oscillatory hormone profiles in the blood, as the alternative would be a flat hormone profile due to the asynchronous action of one million islets. However, this assumption must be reconsidered because recent results raise the possibility that islets are not functionally identical [12].

Here, we model the intra- and inter-islet network, and examine the glucose regulation by the hierarchical islet network. The native intra-islet network is composed of anti-symmetric interactions between α, β, and δ cells. We consider all possible alternatives to the native network motif, and examine their phase coordination between insulin and glucagon pules and between islets and their glucose regulation. Then, we demonstrate that the native motif is one of the most effective motifs that stably regulate glucose levels with minimal hormone consumption, and flexibly control inter-islet synchronization depending on glucose levels.

**Intra-islet network and glucose homeostasis**

To probe how unique the native intra-islet network is for controlling inter-islet synchronization and stably regulating glucose levels, we formulate the glucose-regulation system on the basis of four key observations:

(i) Islet cells produce hormone pulses depending on glucose levels [13];

(ii) Islet cells interact with each other with a special symmetry [9];

(iii) Insulin and glucagon regulate glucose levels [14];

(iv) Oscillatory glucose levels entrain islets to produce synchronous hormone pulses [15-17].

The spontaneous hormone pulses and their interactions (i, ii) can be simply described by a coupled oscillator model [18]. Here, we extend the model for multiple islets and explicitly consider the glucose regulation and entrainment (iii, iv). By considering glucose as a dynamic variable in the whole system, glucose stimulates islets, while islets regulate glucose.

The spontaneous hormone pulses or oscillations of α, β, and δ cells can be described by the amplitude and phase of a generic oscillator [19]:

$$\dot{r}_{n\sigma} = \tau_r^{-1}\left[f_\sigma(G) - r_{n\sigma}^2\right] r_{n\sigma}, \quad (1)$$

$$\dot{\theta}_{n\sigma} = \omega_{n\sigma} - g_\sigma(G)\cos\theta_{n\sigma} \quad (2)$$

for the $\sigma \in \{\alpha,\beta,\delta\}$ cell type in the *n*th islet. The differential equations generate oscillations with a stationary amplitude of $r_{n\sigma} = \sqrt{f_\sigma}$ and an intrinsic phase velocity of $\omega_{n\sigma}$. We focus on *slow* hormone oscillation with a period of $2\pi / \omega_{n\sigma} = 5 \pm 1$ minutes [2]. A characteristic time constant, $\tau_r = 1$ min, is introduced to match the time scale for amplitude and phase dynamics. Given the amplitude and phase, the hormone secretion from the $\sigma$ cell in the *n*th islet is defined as $H_{n\sigma} \equiv r_{n\sigma}(\cos\theta_{n\sigma} + 1)$, where the phase $\theta_{n\sigma} = 0$ and $\pi$ represent maximal and basal secretion, respectively. The cosine has been shifted by unity to prevent negativity of hormone secretion. The amplitude modulation, $f_\sigma(G)$, depends on the glucose concentration, $G$. Briefly $f_\alpha$ and $f_\beta$ are decreasing and increasing functions of $G$, respectively, as α cells secrete glucagon at low glucose levels, while β cells secrete insulin at high glucose levels [20]. Furthermore, the phase modulation, $g_\sigma(G)$, accounts for the fact that hormone pulses are not pure sine waves. They modulate the duration of active and silent phases depending on the glucose concentration. Islet $Ca^{2+}$ oscillations imply that insulin pulses have longer active phases at higher glucose levels [21-23] (See Section 1 in Supplementary Information for the detailed description for the model including $f_\sigma(G)$ and $g_\sigma(G)$).

Each islet responds to a global glucose concentration, $G$, and secretes hormones accordingly (Fig. 1A). The total glucagon, $N \cdot H_\alpha = \sum_{n=1}^{N} H_{n\alpha}$, from *N* islets increases $G$, while the total insulin, $N \cdot H_\beta = \sum_{n=1}^{N} H_{n\beta}$, decreases $G$. Unlike the positive glucose flux associated with glucagon, the negative glucose flux associated with insulin is proportional to the present glucose concentration. The following equation summarizes glucose regulation:

$$\dot{G} = \lambda N(G_0 H_\alpha - G H_\beta) + I(t), \quad (3)$$

where $\lambda$ represents the effectiveness of hormone action for glucose regulation, and $I(t)$ represents external glucose inputs reflecting food intake (positive input) or exogenous insulin injection (negative input). Here, we introduce a constant parameter ($G_0 = 7$ mM) to match the scale for glucagon and insulin action. In the absence of glucose inputs ($I = 0$), a normal glucose concentration is approximated by $G \approx G_0$ because the glucagon and insulin amplitudes balance ($r_{n\alpha} \approx r_{n\beta}$) at $G_0$ and vanish the glucose regulation ($\dot{G} = 0$). Given positive/negative glucose inputs, the islet system produces more insulin/glucagon to less perturb the normal glucose concentration (Fig. 1B). Here, islets do not interact directly with each other, but the total hormonal secretions modulate glucose, which in turn affects each islet. For example, β cells secrete insulin, which reduces glucose and thereby inhibits β cells and stimulates α cells. This is a kind of mean field model in which individual islets interact indirectly though the mean field of global glucose. Consequently, spontaneous hormone oscillations lead to oscillations of glucose. The amplitude variation in the glucose concentration can contribute to entraining islets to exhibit similar phases [24,25]. *Glucose entrainment* is one mechanistic explanation for inter-islet synchronization with the mechanism mediated by the neural pacemaker within the pancreas [13].

Now, we consider the intra-islet interaction between α, β, and δ cells and their roles for regulating glucose homeostasis. First, using a complex variable, $Z_{n\sigma} \equiv r_{n\sigma} e^{i\theta_{n\sigma}}$, and its complex conjugate, $Z^*_{n\sigma}$, we combine the amplitude and phase dynamics in Eqs. (1) and (2): $\dot{Z}_{n\sigma} = J_{n\sigma} Z_{n\sigma}$, where $J_{n\sigma} \equiv f_\sigma - Z_{n\sigma} Z^*_{n\sigma} + i\omega_{n\sigma} - ig_\sigma (Z_{n\sigma} + Z^*_{n\sigma})(4 Z_{n\sigma} Z^*_{n\sigma})^{-1/2}$. Then, in the presence of the intra-islet interaction, the islet model is written as follows:

$$\dot{Z}_{n\sigma} = J_{n\sigma} Z_{n\sigma} + K \sum_{\sigma'} A_{\sigma\sigma'} Z_{n\sigma'}, \qquad (4)$$

where $K$ represents the interaction strength, and the adjacency matrix, $A_{\sigma\sigma'}$, represents the interaction signs from $\sigma'$ cell to $\sigma$ cell within each islet. Positive/negative interaction leads the amplitude and phase of $\sigma$ cell to be positively/negatively correlated to the amplitude and phase of $\sigma'$ cell. For example, positive interaction from $\sigma'$ cell stimulates neighboring $\sigma$ cell to produce larger oscillation and pull its phase to be in phase with $\sigma'$ cell. Ignoring self-interactions ( $A_{\sigma\sigma} = 0$ ), a total of 729 (=3[6]) networks are possible with either positive, no, or negative interactions for each link (Fig. 1C). We then conducted glucose regulation simulations with various glucose stimuli, $I$ , and obtained the corresponding glucose level and glucagon and insulin secretions for all possible networks (Fig. 1D). All of the networks could reasonably control the glucose stimuli by balancing the antagonistic hormones glucagon and insulin. Interestingly, native islet network 121212 emerged as one of the most efficient networks requiring minimal hormone secretion to achieve glucose regulation (See Section 2 in the Supplementary Information for the parameter independence of our conclusion).

**Effective networks for minimal hormone consumption**

In glucose regulation balanced by glucagon and insulin, some wasteful cosecretion of the antagonistic hormones is possible. Thus, we examined the hormone consumption, $H_\alpha + H_\beta$, of the 729 networks under normal ( $I = 0$ ) and high ( $I = 2G_0$ ) glucose conditions for 300 min and 200 min, respectively (See Section 3 in the Supplementary Information for the total hormone consumption including somatostatin and temporal behaviors of three hormones). After removing initial equilibration periods, we calculated average hormone consumptions for the last $T = 100$ min: $H_{\text{normal}} = T^{-1}\int_{200}^{200+T}[H_\alpha(t) + H_\beta(t)]dt$ and $H_{\text{high}} = T^{-1}\int_{400}^{400+T}[H_\alpha(t) + H_\beta(t)]dt$ (Fig. 2A). As α and β cells are key components of glucose regulation, first we focused on the mutual interaction between these cells by comparing four groups of networks: (i) mutual activation networks (11xxxx); (ii) mutual inhibition networks (22xxxx); (iii) native asymmetric networks (12xxxx); and (iv) inverse asymmetric networks (21xxxx). We expected that the mutual inhibition networks could effectively prevent the cosecretion of insulin and glucagon. However, the mutual activation/inhibition networks generally consume larger amounts of hormones because the mutual interaction has the same sign and construct positive feedback loops (α→β →α and β→α→β) regardless of the signs (positive/negative). The mutual activation/inhibition networks have positive coupling terms, $A_{\alpha\beta}\cos(\theta_{n\beta} - \theta_{n\alpha})$ and $A_{\beta\alpha}\cos(\theta_{n\alpha} - \theta_{n\beta})$, in the amplitude component of Eq. (4), which increase both amplitudes, $r_{n\alpha}$ and $r_{n\beta}$. In contrast, the asymmetric networks consume smaller amounts of hormones than the symmetric networks. In particular, 12xxxx networks consume lower levels of hormones than 21xxxx networks at high glucose because α/β cells are successfully suppressed/enhanced to decrease glucose levels.

Next, we checked the robustness of glucose regulation by quantifying the fluctuations of the glucose concentration: $\delta G^2 = T^{-1}\int_{t_0}^{t_0+T}[G(t)-\overline{G}]^2 dt$, where $\overline{G} = T^{-1}\int_{t_0}^{t_0+T} G(t)dt$. The corresponding fluctuations under normal and high glucose conditions were calculated with $t_0 = 200$ min and $t_0 = 400$ min, respectively (Fig. 2B). The amplitude variations in the glucose concentration can entrain islets to produce synchronized hormone pulses. Then, the large variations in synchronized hormones from the entrained islets may cause larger-amplitude variations in the glucose concentration. Thus, a strong correlation is expected between glucose fluctuations (Fig. 2B) and inter-islet synchronization (Fig. 2C). Here, to quantify the phase synchronization between islets, we adopted the usual synchronization index for oscillators, $\rho_\sigma e^{i\Theta_\sigma} = N^{-1}\sum_{n=1}^{N} e^{i\theta_{n\sigma}}$, where $\rho_\sigma$ measures the degree of synchronization between $\sigma$ cells located in different islets (0/1 for perfect desynchronization/synchronization), and $\Theta_\sigma$ captures their average phase at a certain time. In particular, we used their time-averaged index, $\rho = (3T)^{-1}\int_{t_0}^{t_0+T}[\rho_\alpha(t)+\rho_\alpha(t)+\rho_\alpha(t)]dt$ with $t_0 = 200$ min for $\rho_{\text{normal}}$ and $t_0 = 400$ min for $\rho_{\text{high}}$. In general, 11xxxx networks show small glucose fluctuations and low inter-islet synchronization, while 22xxxx networks show large glucose fluctuations and high inter-islet synchronization. Mutual activation networks exhibit *frustration* for the coordination between α and β cells: (internal) mutual positive interactions lead them to display in-phase coordination, while (external) glucose leads them to present out-of-phase coordination. The frustration hinders different islets from generating coherent behavior under glucose entrainment. On the contrary, mutual inhibition networks do not experience frustration, and every islet is easily entrained to external glucose with out-of-phase coordination between α and β cells.

This strong correlation between glucose fluctuations and inter-islet synchronization is not always found in networks 12xxxx and 21xxxx. At normal glucose concentrations, these networks show small glucose fluctuations and low inter-islet synchronization. However, at high glucose levels, the asymmetric networks show a wide range of glucose fluctuations and inter-islet synchronization. Some networks, including the native islet network 121212, exhibit small glucose fluctuations with high inter-islet synchronization, which seems counterintuitive because the large variations in coherent hormone pulses from synchronized islets can lead to large fluctuations of glucose concentrations governed by Eq. (3). Furthermore, the out-of-phase coordination between $H_\alpha$ and $H_\beta$ at high glucose levels can amplify glucose fluctuations by alternatively increasing and decreasing glucose concentrations. The network 121212 generates synchronous hormone pulses of $H_{n\alpha}$ and $H_{n\beta}$, and has the out-of-phase coordination between their average glucagon $H_\alpha$ and insulin $H_\beta$. Nevertheless, it does not induce large glucose fluctuations because the combined phase of the glucose and insulin action ($G \cdot H_\beta$) is not out of phase with $H_\alpha$. The combined phase, not the pure insulin phase, regulates $\dot{G}$ in Eq. (3). The effective phase coordination between glucose and insulin allows $\dot{G}$ to remain negligibly small in stationary states. Thus, network 121212 can achieve coherent hormone pulses of synchronized islets but maintain small glucose fluctuations at high glucose levels.

Effective homeostatic networks should tightly regulate the glucose concentration with small fluctuations by consuming minimal amounts of hormones. Among a total of 729 possible networks, we sorted out the effective networks that consume smaller amounts of hormones than network 000000 (no interaction between islet cells) and show small glucose fluctuations ($\delta G / G$ < 0.1) (Fig. 2D). The applied criteria identified ten effective networks (Fig. 2E and Table 1). Indeed, these networks are subsets or transformations of network 121212 or 122112. In other words, if one removes some links or changes cell names from 121212 or 122112, the remaining eight networks can be obtained. We noted that the asymmetric interactions between islet cells are prevalent in the ten effective networks, and mutual activation/inhibition never occurs between any pair of cells. In particular, the asymmetric interaction between β and δ cells are well conserved: nine effective networks have the xxxx12 type in which β cells activate δ cells while δ cells suppress β cells. The ten effective networks could stably regulate glucose with minimal consumption of two antagonistic hormones, insulin and glucagon. Their somatostatin consumption was also relatively minimal (Table S2). Moreover, they showed controllable inter-islet synchronization with low/high synchronization index at normal/high glucose (Table 1). Among them, the fully connected networks 121212, 121221, and 122112 showed higher inter-islet synchronization at high glucose. Here, network 121221 is topologically equivalent to network 121212. The name switch between β and δ cells can transform network 121221 to network 121212. On the other hand, networks 121212 and 122112 are topologically distinct. Network 121212 exhibits distinguishable cells: one cell suppresses the other two; one cell enhances the other two; and one cell suppresses one and enhances one. However, network 122112 shows indistinguishable cells: every cell suppresses and enhances the others. In other words, if one hides cell names, one cannot distinguish between cells. Here, network 122212 may require more complex sets of receptors to realize the interaction topology because one cell must affect the other two cells differently. Finally, we have also examined the phase coordination between three hormones for the ten effective networks. Networks 121212 and 120012 clearly reproduced the out-of-phase coordination between insulin and glucagon pulses, and the in-phase coordination between insulin and somatostatin (Fig. S9), as experimentally observed by Hellman et al. [7] In summary, network 121212 outstands in many aspects such as minimal hormone consumption, stable glucose regulation, controllable inter-islet synchronization, and clear phase coordination between three hormones.

**Controlling inter-islet synchronization**

The effective networks display the special feature of controllable inter-islet synchronization depending on glucose conditions. In contrast, although network 000000 (no interaction between islet cells) consumes relatively small amount of hormones among 729 networks, it exhibits a serious problem in that islets are easily entrained to glucose oscillation and synchronized with each other (Fig. 3A). Then, the synchronized hormone secretion amplifies glucose fluctuations. In contrast, networks 121212 and 122112 tightly regulate minimal glucose fluctuations, especially under normal glucose concentrations. Hence, we asked how the effective networks control inter-islet synchronization.

We found that network 121212 and 122112 display different numbers of attractors in the phase dynamics in Eq. (4) under different glucose conditions. They exhibit triple attractors of phase differences ($\theta_{n\alpha} - \theta_{n\beta}, \theta_{n\alpha} - \theta_{n\delta}$) at normal glucose levels, while the distinction between the attractors vanishes at high glucose levels (Fig. 3B and 3C; See Section 4 in the Supplementary Information for the analysis). Thus, different islets sit on different attractors at normal glucose levels. This causes different islets to present different phase coordination between islet cells. This heterogeneous phase coordination hinders islets from being synchronized. With high glucose levels, however, the triple attractors vanish, and islets do not present clearly distinct phase coordination between islet cells. Then, islets become easily synchronized without resistance to glucose entrainment. This scenario is exactly the same under low glucose conditions (See Section 5 in Supplementary Information). Unlike these effective networks, network 000000 shows no distinct attractors, regardless of glucose conditions (Fig. 3D), and glucose oscillations can easily entrain islets to be synchronized. Therefore, we conclude that inter-islet desynchronization at normal glucose levels is an active process that is resistant to glucose entrainment by generating multiple attractors through its dynamics. We also confirmed the controllability of inter-islet synchronization in the population model in which each islet is composed of populations of islet cells (See Section 6 in Supplementary Information).

## Discussion

Pancreatic islets secrete pulsatile hormones to regulate glucose homeostasis. Each islet is composed of α, β, and δ cells that interact with each other with a unique symmetry. We asked how special the islet-cell interaction is for coordinating hormone secretions from multiple islets and effectively regulating glucose homeostasis. Here, we have revealed the link between the intra-islet network and inter-islet synchronization, both of which have been long-standing puzzles in islet biology. First, the anti-symmetric interactions between α, β, and δ cells are unique in being able to prevent the wasteful zero-sum counter-regulatory actions of glucagon and insulin for glucose regulation. More importantly, the intra-islet network contributes to controlling the synchronization between islets in the pancreas, and the phase coordination between hormones. The multiplicity of islets in the pancreas allows signal amplification and suppression, once the coherence/independence of one million islets is controllable. The anti-symmetric regulatory motif embodies the potential of phase modulation in biological oscillations.

Rhythms and their synchronization have been an important issue in living and man-made systems [26-28]. Synchronization is not always desirable. For example, hypersynchronous neuronal activities can cause epileptic seizures in the brain [29]. Therefore, desynchronization and its controllability have been highlighted recently [30-33]. Here, pancreatic islets showed an interesting strategy for controlling synchronization. They exist as multiple micro-organs instead of a single gigantic organ such as liver, lung, and heart. This structural design may be advantageous for generating asynchronous hormone secretion. However, once required, islets can also generate synchronous hormone secretion through the glucose entrainment mechanism. Finally, since the Kuramoto model has been extensively studied for understanding synchronization phenomena [19,26], our study provided an explicit biological origin of the phenomenological model.

We formulated a minimal model for describing the loop of hormone secretion and glucose regulation. The model incorporated four basic observations: (i) glucose-dependent pulsatile hormone secretion; (ii) paracrine interactions between islet cells; (iii) glucose regulation by insulin and glucagon; and (iv) glucose entrainment of multiple islets. This model can provide a platform for simulating glucose regulation in various conditions. We have simulated the glucose regulation under noisy or oscillatory glucose infusions (See Sections 7 and 8 in Supplementary Information). The noisy glucose infusion lead glucose levels more fluctuating, and it did not change our conclusion about the effectiveness of the network 121212. Next, the oscillatory glucose infusion could entrain hormone secretion to follow the glucose rhythm, if the oscillation amplitude of external glucose is sufficiently large. However, the present minimal model was limited to include detailed mechanisms of hormone secretion. In the phenomenological model, hormone oscillations were intrinsically given without explicit consideration of the molecular details for generating the oscillations. Furthermore, we assumed that the amplitude and phase modulations ($f_\sigma$ and $g_\sigma$) were functions of glucose as a first approximation, although they can also depend on glucose change. Islets show an acute insulin secretion when glucose levels quickly change [34-36]. Thus, the minimal model may need to incorporate other relevant observations to serve as a more realistic simulator for glucose regulation.

The pulsatility of insulin secretion has functional advantages. Pulsatile insulin can suppress hepatic glucose production more effectively than constant insulin [37], and it may prevent the desensitization of insulin receptors [38,39]. Furthermore, it has been observed that the insulin pulsatility is diminished in diabetic patients [40]. This can be explained by the diminished pulse generation of β cells under diabetic conditions. However, loss of coordinated hormone secretions from multiple islets can also contribute to the diminished pulsatility of circulating insulin levels in blood. We demonstrated that the intra-islet network was special for controlling the inter-islet synchronization. Furthermore, the intra-islet network governed the phase coordination between glucagon, insulin, and somatostatin pulses (Fig. S9). The out-of-phase coordination between insulin and glucagon pulses should have physiological relevance. Hellman et al. have hypothesized that the inverse relation may enhance the insulin to glucagon ratio for hepatic glucose production [7], and Menge et al. have reported that the inverse relation is disrupted in Type 2 diabetes [8]. Thus, our conclusion suggested that modifications of the local interactions between α, β, and δ cells under diabetic conditions can contribute to the abnormal global coordination of hormone secretions: (i) inter-islet synchronization; and (ii) phase relation between glucagon, insulin, and somatostatin pulses.

**Methods**

The above differential equations were numerically integrated using the Euler method [41] with a sufficiently small time step, $\Delta t = 0.0001$.

**Acknowledgments**

We thank Arthur Sherman, Vipul Periwal, Erik Gylfe, Anders Tengholm, Bo Hellman, Jong Bhak, and Jaekyoung Kim for helpful comments, and the anonymous reviewers for their valuable comments and suggestions to improve the quality of the paper. This research was supported by Basic Science Research Program through the National Research Foundation of Korea (NRF) funded by the Ministry of Education (2016R1D1A1B03932264) and the Max Planck Society, the Korea Ministry of Education, Science and Technology, Gyeongsangbuk-Do and Pohang City (J.J.).

## Author Contributions

D.-H.P., D.-T.H, J.X. and J.J. initiated the project. D.-H.P., T.S., D.-T.H. conducted the numerical simulations. T.S. prepared all the figures. D.-H.P., T.S. and J.J. analyzed the results, and wrote the manuscript. All authors reviewed the manuscript.

## Additional Information

Supplementary information accompanies this paper at http://www.nature.com/srep

Competing financial interests: The authors declare no competing financial interests.

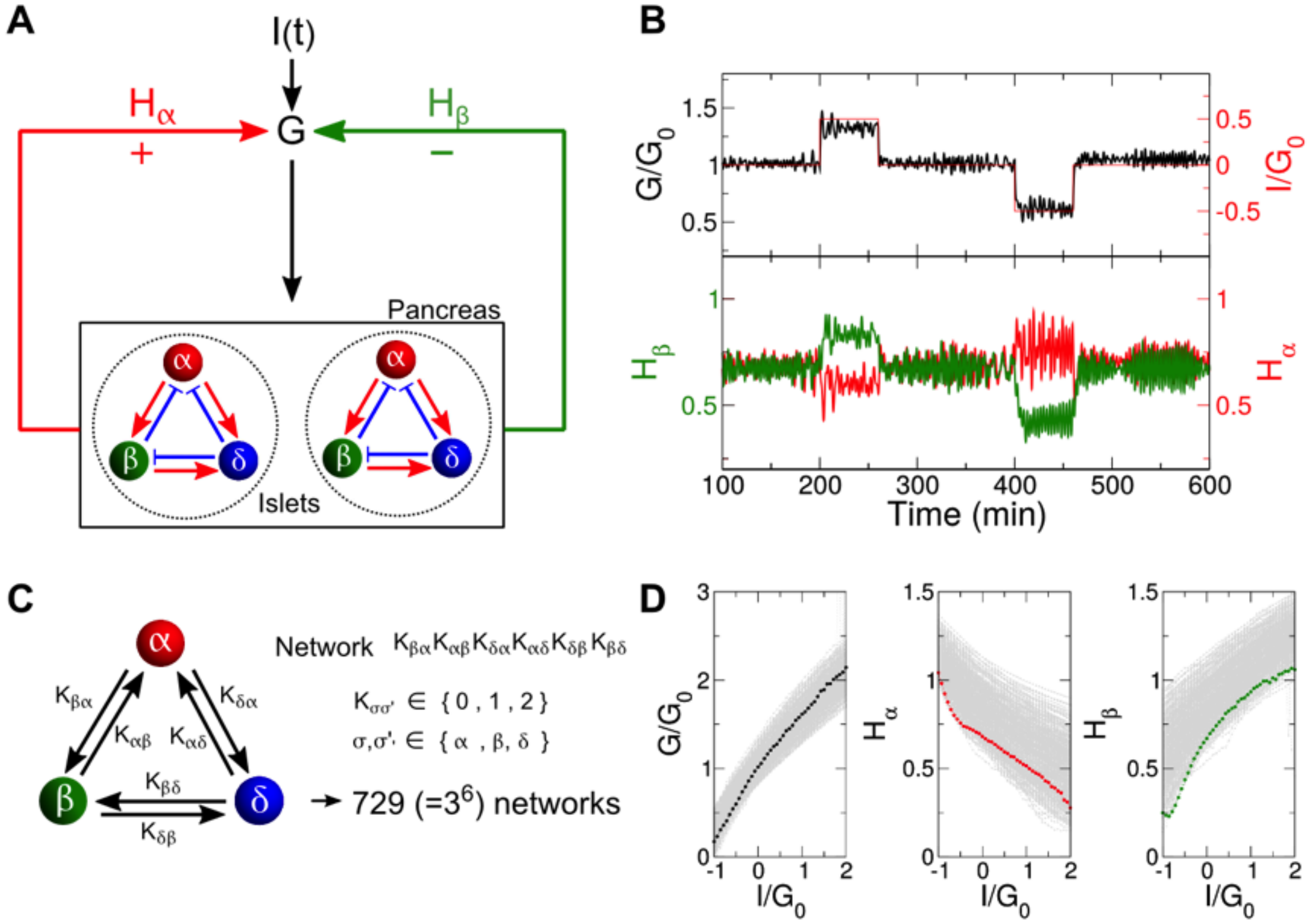

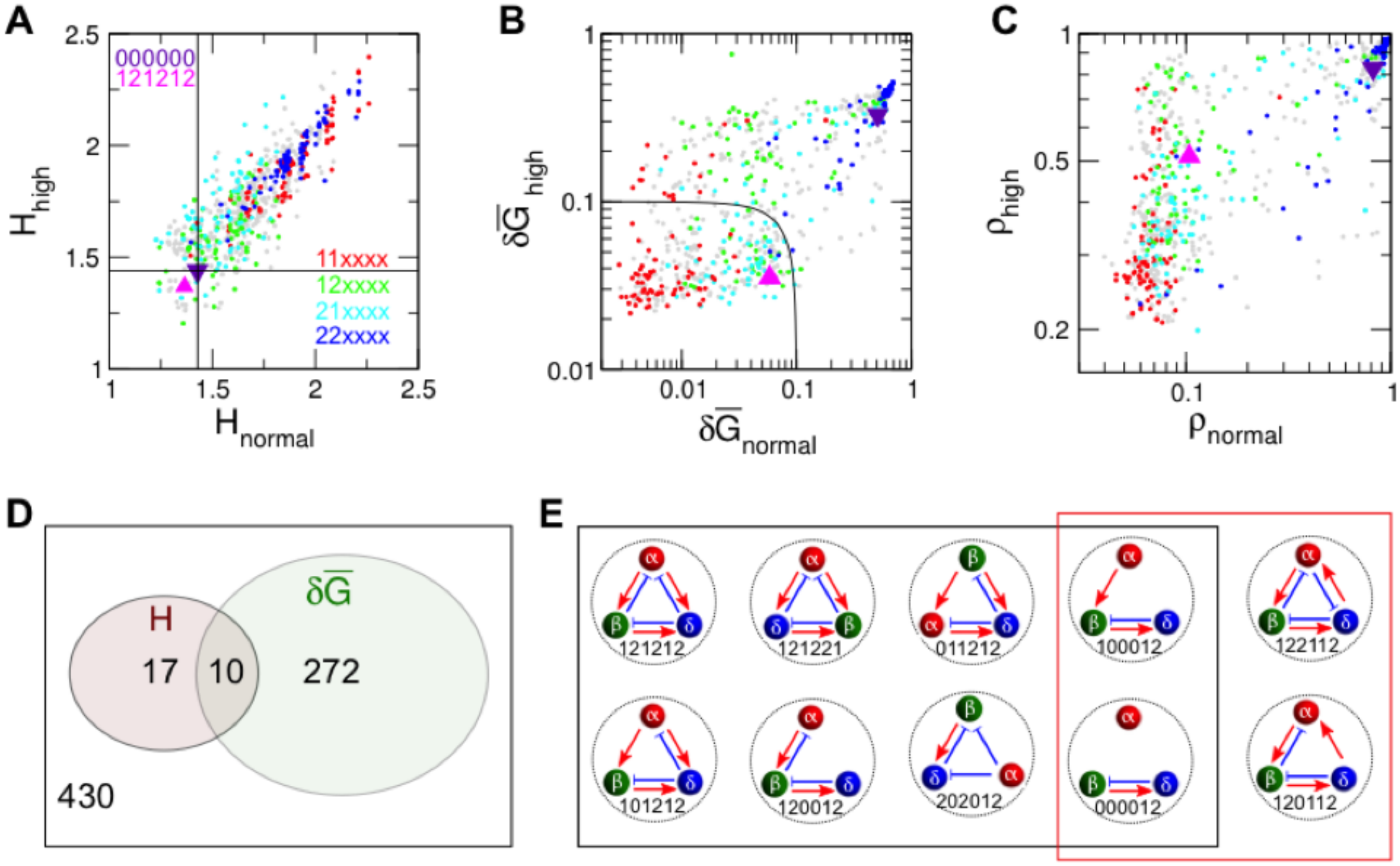

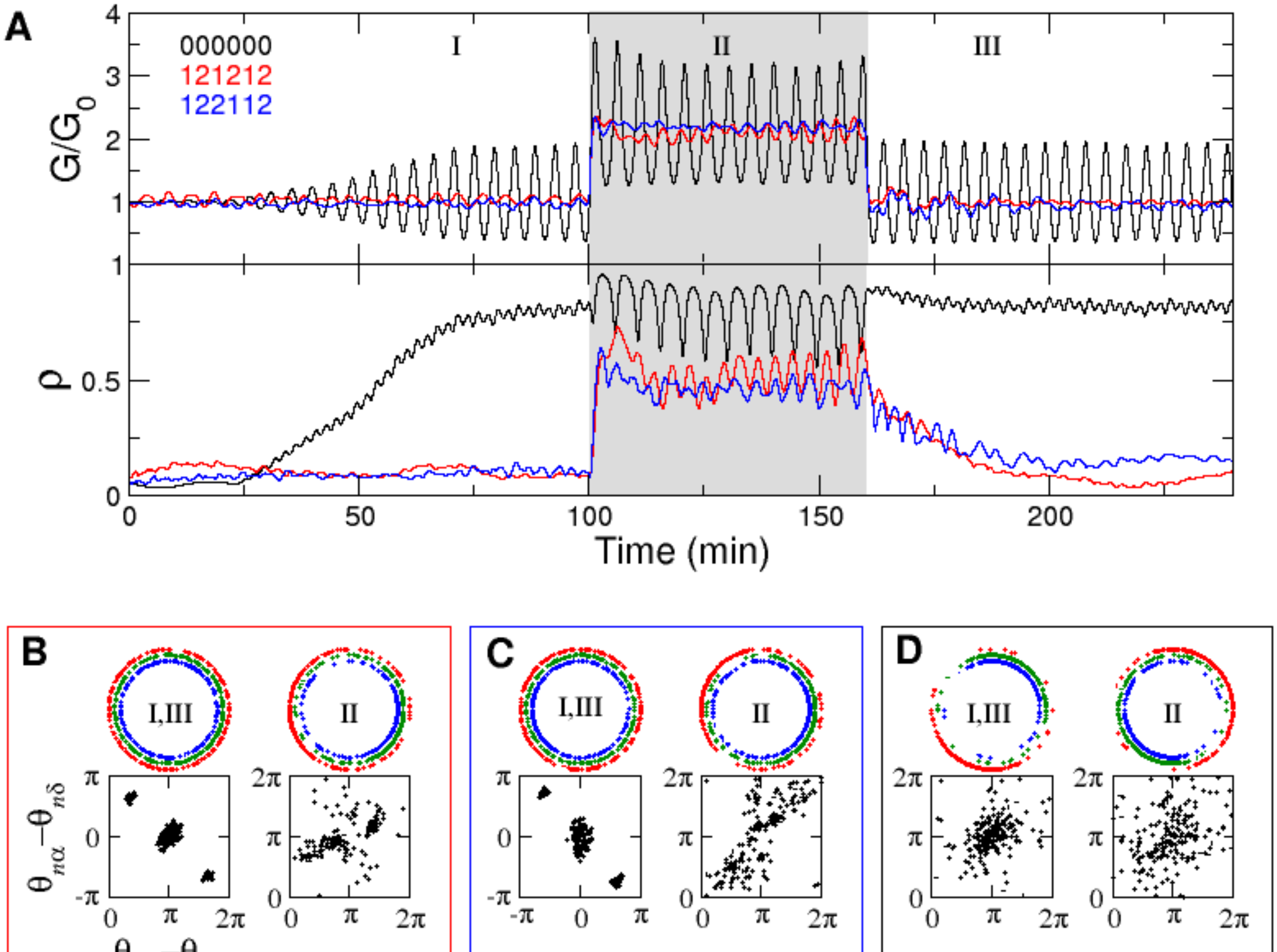

## Figure Legends

**Fig. 1. Glucose regulation and islet cell network.** (A) Schematic diagram of glucose regulation by pancreatic islets. Endocrine α, β, and δ cells in the pancreatic islets monitor and regulate blood glucose concentration, *G*, perturbed by external input, $I$. Glucagon, $H_\alpha$, secreted by α cells increases *G*, while insulin, $H_\beta$, secreted by β cells decreases *G*,. (B) Given the external glucose input, $I(t)$ (red line in upper plot), glucose $G$ (solid black line) is regulated by two counter-regulatory hormones: $H_\alpha$ (red line in lower plot) and $H_\beta$ (green line). Note that $G_0$ is the normal glucose concentration in the absence of an input ($I = 0$). Network 121212 is used. (C) Network annotation for describing the interactions between islet cells. The interaction signs of $A_{\beta\alpha}$, $A_{\alpha\beta}$, $A_{\delta\alpha}$, $A_{\alpha\delta}$, $A_{\delta\beta}$, and $A_{\beta\delta}$ can take zero (0), positive (1), or negative (2) values. Following this notation, network 121212 represents the native network of islet cells. (D) Stationary glucose concentration and the corresponding hormone consumption for various external glucose inputs. Negative/positive inputs, $I$, induce low/high glucose conditions. A total of 729 networks were considered: network 121212 (black, red, green) and the other networks (gray). See Section 2 in the Supplementary Material for the details of the model and standard parameter values.

**Fig. 2. Efficient hormone consumption and glucose regulation.** (A) Hormone consumption, $H \equiv H_\alpha + H_\beta$, of 729 networks under normal ($I = 0$) and high ($I = 2G_0$) glucose conditions: networks 121212 (pink triangle, native interaction between islet cells), 000000 (purple inverted triangle, no interaction), 11xxxx (red circles, mutual activation between α and β cells), 12xxxx (green circles, native asymmetric interaction), 21xxxx (cyan circles, inverse asymmetric interaction), 22xxxx (blue circles, mutual inhibition), and the others (gray circles). Network 000000 is located at the crossing of vertical and horizontal black lines. (B) Temporal fluctuations of glucose, normalized according to the absolute glucose concentration, $\delta\overline{G} \equiv \delta G / G$, under normal/high glucose conditions. Black lines separate the networks that produce small glucose fluctuations ($\delta\overline{G}^2_{\text{normal}} + \delta\overline{G}^2_{\text{high}} < 0.1^2$). (C) Degree of synchronization between islets under normal/high glucose conditions. The time-averaged synchronization index, $\rho = 1/0$, represents perfect synchronization/independence between islets. (D) Ten effective networks satisfy the two criteria of minimal hormone consumption and small glucose fluctuations inside the black lines in (A) and (B). (E) Topologies of the ten effective networks with their annotations. Positive/negative interactions are represented by red/blue bar-headed arrows, respectively. A black rectangle groups subsets and transformations of network 121212, while a red rectangle groups subsets of network 122112. To highlight this, networks are plotted with fixed arrow positions, not cell positions.

**Fig. 3. Controllable inter-islet synchronization and phase coordination between islet cells.** (A) Glucose regulation and inter-islet synchronization for networks 121212 (red), 122112 (blue) and 000000 (black), given the external glucose input ($I = 2G_0$) during $100 < t < 160$. Phase snapshots of α, β, and δ cells under different glucose conditions (regimes I, II, and III) for (B) networks 121212, (C) 122112, and (D) 000000. Upper panel: absolute phases ($\theta_{n\alpha}, \theta_{n\beta}, \theta_{n\delta}$) of α (red), β (green), and δ (blue) cells in the pancreas consisting of 200 islets. Lower panel: phase differences ($\theta_{n\alpha} - \theta_{n\beta}, \theta_{n\alpha} - \theta_{n\delta}$). The axis range was adjusted to show distinct attractors considering 2π periodicity.

**Table 1.** Hormone consumption, glucose fluctuation, and inter-islet synchronization of ten effective networks under normal and high glucose conditions.

| # | Network | $H_{\text{normal}}$ | $H_{\text{high}}$ | $\delta\overline{G}_{\text{normal}}$ | $\delta\overline{G}_{\text{high}}$ | $\rho_{\text{normal}}$ | $\rho_{\text{high}}$ |
|---|---|---|---|---|---|---|---|
| 1 | 000012 | 1.357 | 1.288 | 0.048 | 0.052 | 0.089 | 0.319 |
| 2 | 011212 | 1.292 | 1.351 | 0.012 | 0.048 | 0.065 | 0.335 |
| 3 | 202012 | 1.329 | 1.299 | 0.039 | 0.040 | 0.116 | 0.309 |
| 4 | 100012 | 1.368 | 1.261 | 0.011 | 0.039 | 0.063 | 0.321 |
| 5 | 101212 | 1.295 | 1.395 | 0.015 | 0.034 | 0.082 | 0.343 |
| 6 | 120012 | 1.244 | 1.383 | 0.041 | 0.041 | 0.090 | 0.373 |
| 7 | 120112 | 1.418 | 1.311 | 0.011 | 0.032 | 0.073 | 0.419 |
| 8 | 122112 | 1.359 | 1.204 | 0.076 | 0.031 | 0.144 | 0.466 |
| 9 | 121221 | 1.360 | 1.378 | 0.085 | 0.038 | 0.171 | 0.475 |
| 10 | 121212 | 1.365 | 1.369 | 0.058 | 0.035 | 0.104 | 0.513 |

Supplementary Information

# A Local Counter-Regulatory Motif Modulates the Global Phase of Hormonal Oscillations

Dong-Ho Park[1†], Taegeun Song[1†], Danh-Tai Hoang[1,2,3†], Jin Xu[1,4], and Junghyo Jo[1,4*]

[1]Asia Pacific Center for Theoretical Physics, Pohang, Gyeongbuk 37673, Korea

[2]Laboratory of Biological Modeling, National Institute of Diabetes and Digestive and Kidney Diseases, National Institutes of Health, Bethesda, Maryland 20892, United States of America

[3]Department of Natural Sciences, Quang Binh University, Dong Hoi, Quang Binh 510000, Vietnam

[4]Department of Physics, Pohang University of Science and Technology, Pohang, Gyeongbuk 37673, Korea

*Correspondence: jojunghyo@apctp.org

†These authors contributed equally to this work.

## 1. Islet model and glucose regulation

Here, we provide a complete description of hormone secretion and glucose regulation in Eqs. (3) and (4):

$$\dot{r}_{n\sigma} = \tau_r^{-1}\left[f_\sigma(G) - r_{n\sigma}^2\right] r_{n\sigma} + K \sum_{\sigma'} A_{\sigma\sigma'} r_{n\sigma'} \cos(\theta_{n\sigma'} - \theta_{n\sigma}), \quad \text{(S1)}$$

$$\dot{\theta}_{n\sigma} = \omega_{n\sigma} - g_{n\sigma}(G)\cos\theta_{n\sigma} + K \sum_{\sigma'} A_{\sigma\sigma'} \frac{r_{n\sigma'}}{r_{n\sigma}} \sin(\theta_{n\sigma'} - \theta_{n\sigma}), \quad \text{(S2)}$$

$$\dot{G} = \lambda\left[G_0 \sum_{n=1}^{N} r_{n\alpha}(1+\cos\theta_{n\alpha}) - G \sum_{n=1}^{N} r_{n\beta}(1+\cos\theta_{n\beta})\right] + I(t). \quad \text{(S3)}$$

The islet model considers amplitude and phase modulations according to the glucose concentration. First, amplitude modulation controls the glucose-dependent hormone secretion of α, β, and δ cells (Fig. S1):

$$f_\alpha(G) = \frac{1}{2}\left[1 - \tanh\left(\frac{G - G_0}{5}\right)\right], \qquad \text{(S3)}$$

$$f_\beta(G) = \frac{1}{2}\left[1 + \tanh\left(\frac{G - G_0}{5}\right)\right], \qquad \text{(S4)}$$

$$f_\delta(G) = \frac{a_\delta}{2}\left[1 + \tanh\left(\frac{G - G_0 + \Delta G_0}{5}\right)\right], \qquad \text{(S5)}$$

where $G_0 = 7$ mM, $\Delta G_0 = 2$ mM, and $a_\delta = 0.5$. The amplitude modulations are based on the observed glucose dose response of insulin, glucagon, and somatostatin secretions (*1*). Note that the U-shaped glucagon response under extremely high glucose (>20 mM) concentrations is not considered here. The somatostatin secretion of δ cells exhibits a lower glucose threshold than the insulin secretion. In addition, we parameterize the lower fraction of δ cells ($a_\delta < 1$) compared with two major populations of α and β cells.

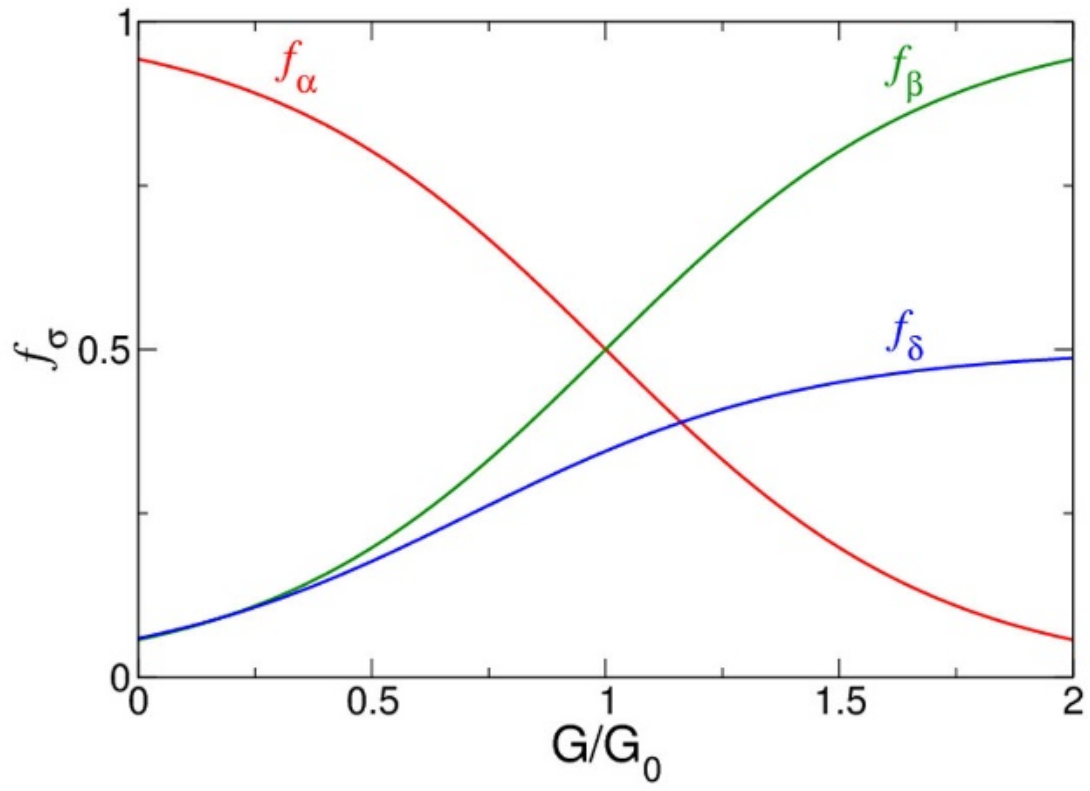

**Fig. S1. Glucose-dependent amplitude modulation.** α cells ($f_\alpha$, red), β cells ($f_\beta$, green), and δ cells ($f_\delta$, blue).

Second, phase modulation controls the duration of active/silent phases in the hormone pulses:

$$g_\alpha(G) = \mu(G - G_0), \qquad \text{(S6)}$$

$$g_\beta(G) = \mu(G_0 - G), \qquad \text{(S7)}$$

$$g_{\delta}(G) = \mu(G_0 - G) . \qquad \text{(S8)}$$

These phase modulations lead α cells to exhibit longer active phases at low glucose levels ( $G < G_0$ ) and β and δ cells to exhibit longer active phases at high glucose levels ( $G > G_0$ ). The glucose-dependent $Ca^{2+}$ oscillations of islets showed that their active phases increase as glucose concentration increases (*2*). Since β cells are dominant (~80%) in islets, the result may support the phase modulation of β cells as shown in Fig. S2. However, direct experimental evidence still lacks for the phase modulations of α and δ cells due to the difficulty of cell identification. Here, we assumed that the phase modulation of α cells is opposite to β cells, while the phase modulation of δ cells is the same with β cells like their amplitude modulations. These descriptions complete the amplitude and phase dynamics in Eqs. (S1) and (S2).

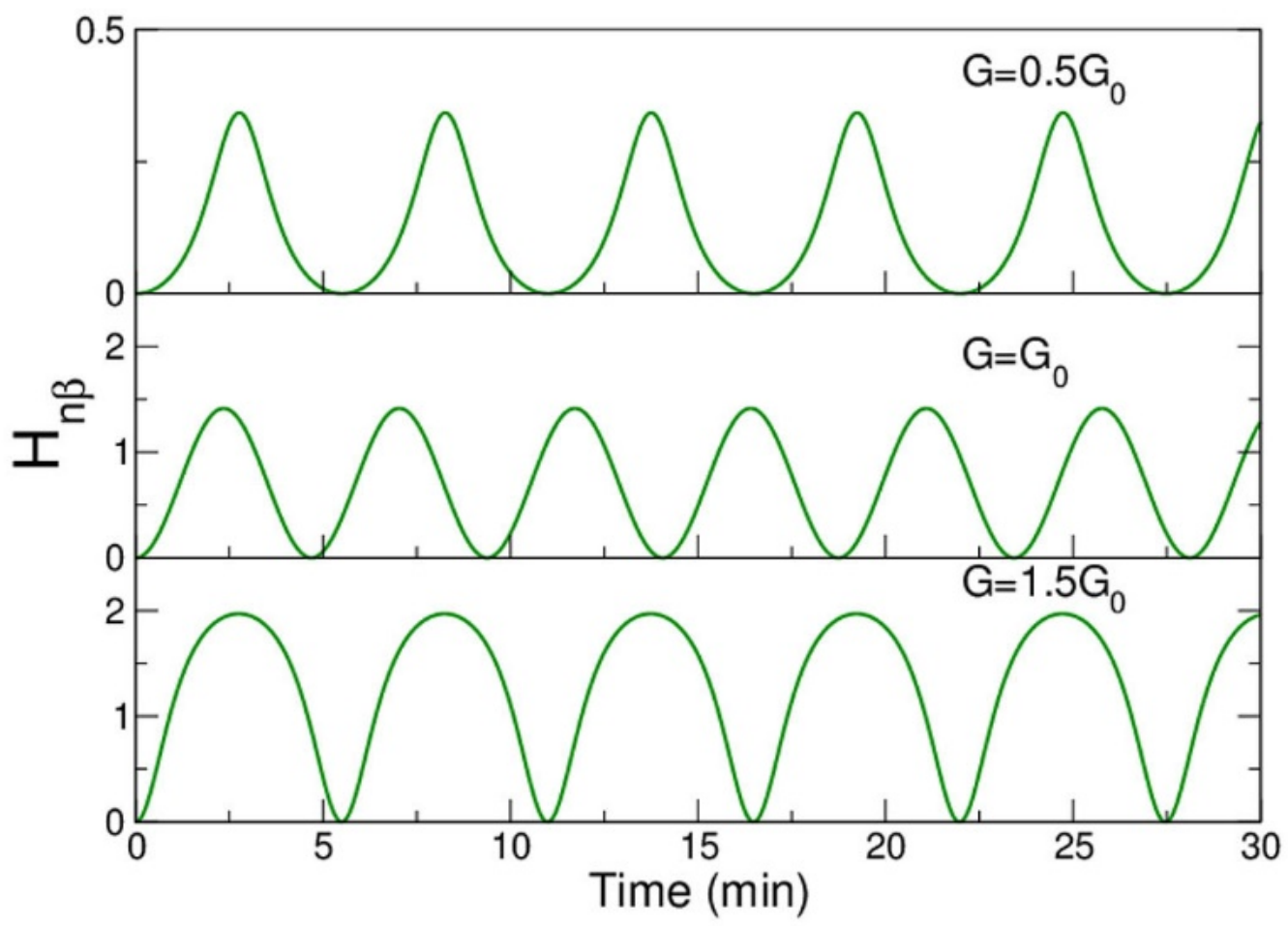

**Fig. S2. Phase modulation.** Glucose-dependent pulse shapes of insulin from a single β cell for various external glucose inputs.

One technical note is that because the negative amplitude $r_{n\sigma} = -\sqrt{f_{\sigma}}$ is another approximate solution of Eq. (S1), we avoid the negative amplitude through transformation as $r_{n\sigma} \rightarrow -r_{n\sigma}$ and $\theta_{n\sigma} \rightarrow \theta_{n\sigma} + \pi$ whenever a negative amplitude is confronted in simulations.

Each cell exhibits heterogeneous intrinsic phase velocities, $\omega_{n\sigma}$ , of which periods follow a Gaussian distribution with a mean of 5 min and a standard deviation of 1 min. Table S1 summarizes our standard parameter values. The hormone and glucose dynamics was not sensitive to initial conditions given a large number of total islets ($N$=200). In this study, we used $r_{n\sigma}(0)\in\{0.25,0.75\}$, $\theta_{n\sigma}(0)\in\{0,2\pi\}$, and $G(0) = 7$ mM.

**Table S1. Standard parameter values.**

| Symbol | Values | Parameter |
| --- | --- | --- |
| $2\pi/\omega_{n\sigma}$ | 5 ± 1 min | Intrinsic periods of islet cells |
| $\tau_r$ | 1 min | Characteristic time for amplitude change |
| $K$ | 0.4 $min^{-1}$ | Interaction rate between islet cells |
| $\mu$ | 0.1 $min^{-1}$ | Phase modulation rate |
| $\lambda$ | 1.0 $min^{-1}$ | Hormone effectiveness rate |
| $G_0$ | 7 mM | Normal glucose concentration |
| $\Delta G_0$ | 2 mM | Threshold shift of δ cells |
| $a_\delta$ | 0.5 | Amplitude reduction of δ cells |
| $N$ | 200 | Total islet number |

## 2. Parameter dependence of the islet model

Although we formulated the islet model constrained by experimental observations, exact information on parameter values is still lacking. Thus, we check the dependences of our conclusions.

(1) Interaction rate between islet cells, $K$

A too small $K$ cannot distinguish the topological difference between islet-cell networks, while a too strong $K$ induces strong nonlinear effects on the phase dynamics in Eq. (S2) and perturbs the principal spontaneous oscillations governed by the intrinsic frequency $\omega_{n\sigma}$. If $K$ is sufficiently large ($K > 0.4$), the native network 121212 shows robust characteristics of small hormone consumption and small glucose fluctuations (Fig. S3).

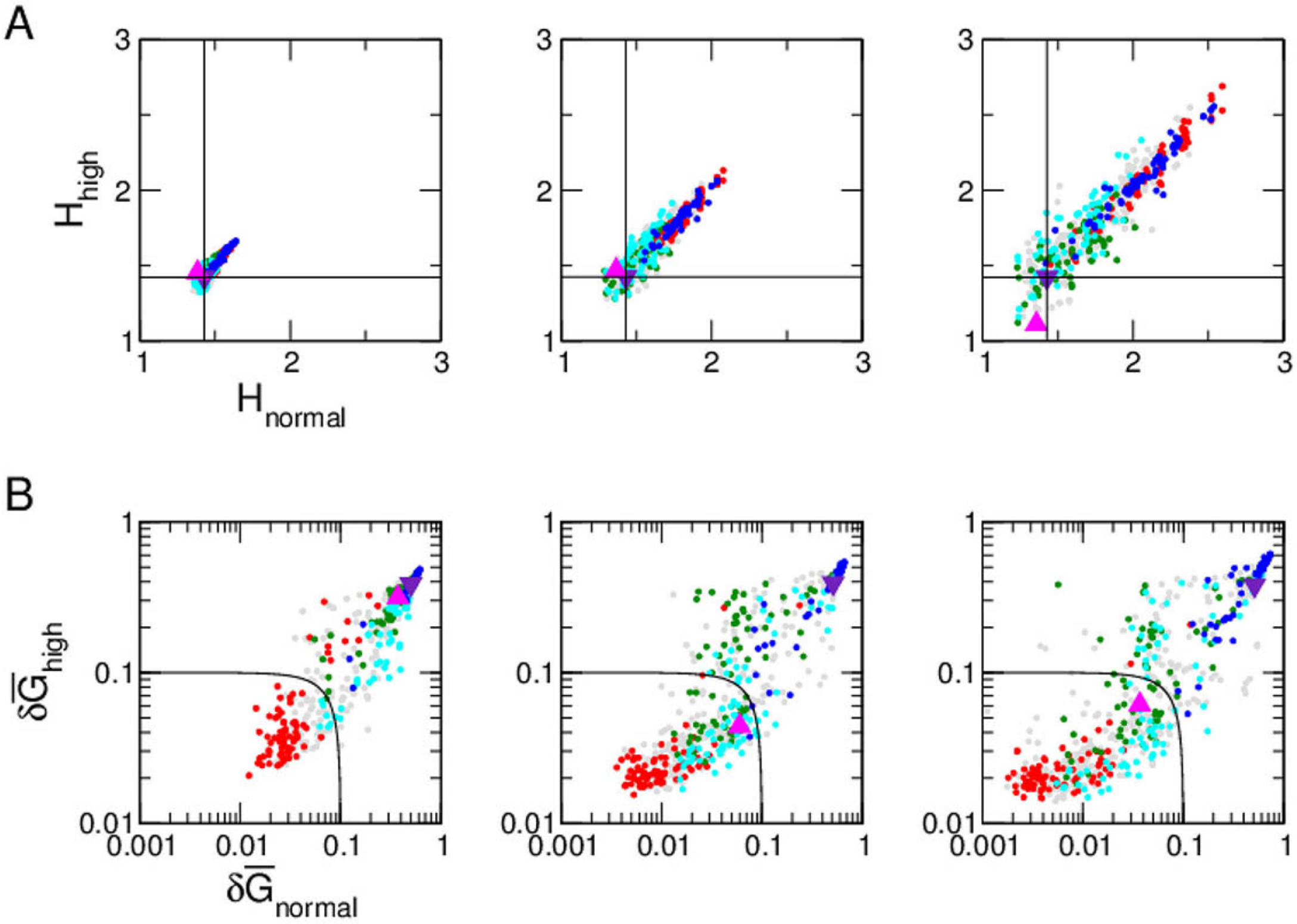

**Fig. S3. Intra-islet coupling strength.** Hormone consumption of 729 networks under normal ($I = 0$) and high ($I = 0.5G_0$) glucose conditions (A) and the temporal fluctuations of glucose (B) under various interaction rates, $K$ = 0.1, 0.3, and 0.6, from left to right. The colors and notations are the same as in Fig. 2.

(2) Phase modulation rate, $\mu$

Phase modulation through a pinning term in Eq. (S2) captures the glucose-dependent shapes of hormone pulses. Thus, the pinning term represents the response of islets to glucose perturbations. The oscillatory changes in glucose are then able to entrain islets to become synchronized. Here too weak pinning cannot induce inter-islet synchronization, while too strong pinning ($|\mu| > \omega_{n\sigma}$) will stop the oscillation of islet cells due to being stuck at $\theta_{n\sigma} = \cos^{-1}(\omega_{n\sigma} / \mu)$. At a lower pinning strength, network 000000 (no interaction between islet cells) starts to show inter-islet synchronization compared with network 121212 (Fig. S4). In addition, given the pinning strength ($0.07 < \mu < 0.2$), network 121212 generates different degrees of inter-islet synchronization for different external glucose inputs, $I$. The mean glucose concentration, $\overline{G}$, is independent of the pinning strength, but glucose fluctuations are highly correlated with inter-islet synchronization.

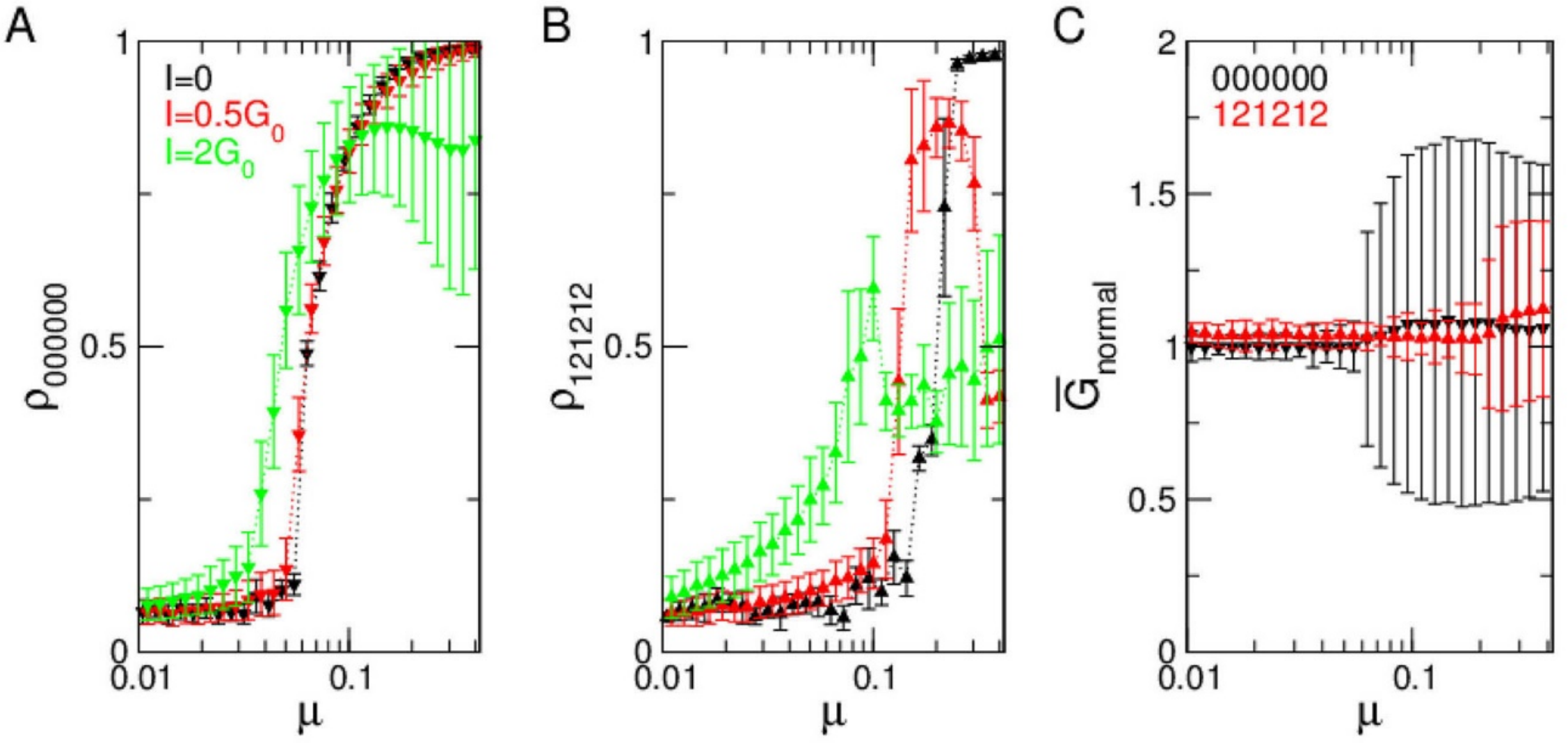

**Fig. S4. Pinning strength.** Inter-islet synchronization index, $\rho$, for networks (A) 000000 and (B) 121212 under external glucose inputs of $I = 0$ (black), $I = 0.5G_0$ (red), and $I = 2G_0$ (green). (C) Mean glucose concentration, $\overline{G}$, for networks 000000 (black) and 121212 (red) under normal glucose conditions ($I = 0$). Error bars represent the standard deviation.

(3) Amplitude reduction of δ cells, $a_\delta$

Considering the minority of δ-cell populations, the value of $a_\delta$ is expected to be $0 < a_\delta < 1$.

Glucose, hormone, and inter-islet synchronization profiles are not highly dependent on the variation of this parameter (Fig. S5). Two extreme networks, 120000 and 000202, correspond to ignorance ( $a_\delta = 0$ ) and emphasis ( $a_\delta > 1$ ) of the interactions of δ cells.

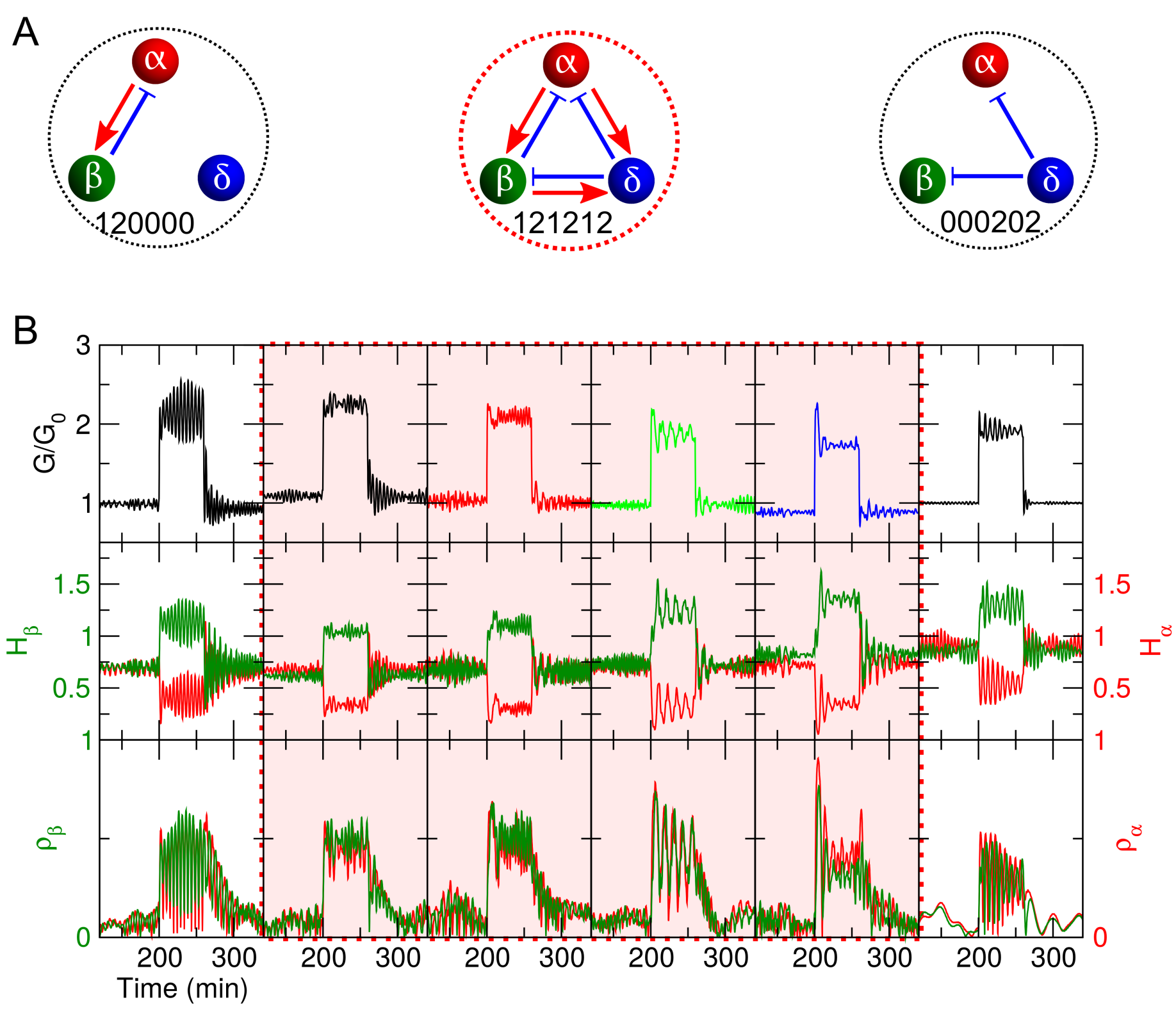

**Fig. S5. Effect of δ-cell amplitude.** (A) Two extreme networks, 120000 and 000202, ignoring and emphasizing the interactions of δ cells, respectively, compared with the native islet network, 121212. (B) Time traces of glucose, hormones, and inter-islet synchronization indices under a glucose stimulus of $I = 2G_0$ for 200 < Time < 260. The red-dotted box represents the results for network 121212 for $a_\delta$= 0.1, 0.5, 1, and 2 from left to right. The leftmost column is the result of network 120000, and the rightmost column is the result of network 000202.

(4) Hormone effectiveness rate, $\lambda$

A too small $\lambda$ fails to effectively regulate glucose in Eq. (2), while a too large $\lambda$ diminishes the relative contribution of external glucose inputs, $I$, in Eq. (2) (Fig. S6).

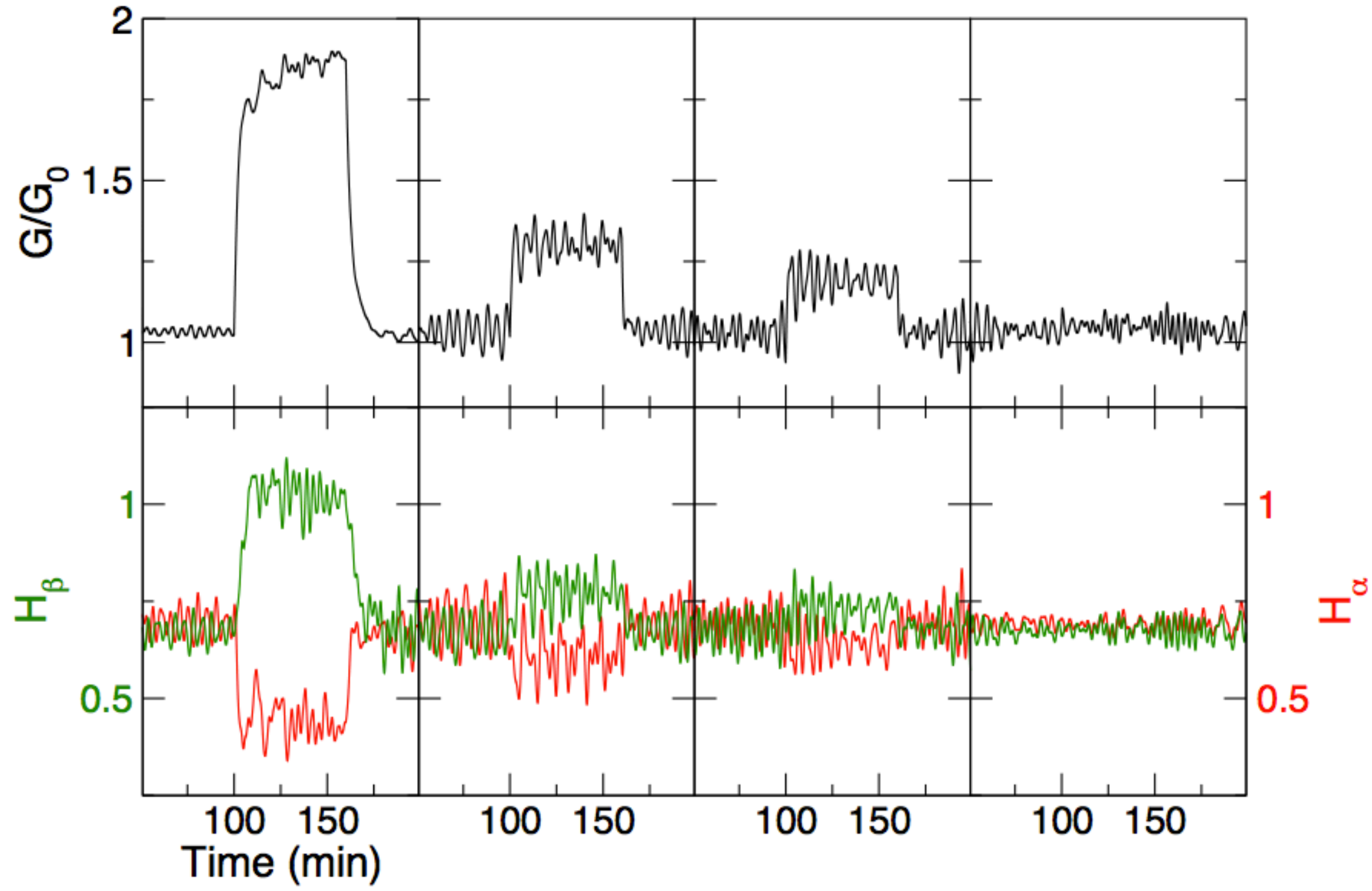

**Fig. S6. Hormone effectiveness rate for glucose regulation.** Time traces of glucose concentrations (upper) and hormones (lower) with various hormone effectiveness rates, $\lambda N$ = 0.2, 0.6, 1, and 10 min$^{-1}$, from left to right under a glucose stimulus of $I = 0.5G_0$ for 100 < Time < 160. For the simulation, network 121212 was used with total $N$=200 islets.

(5) Total islet number, $N$

As $N$ increases, the fluctuation of glucose regulation decreases (Fig. S7). In addition, the initial condition dependence of the nonlinear dynamics is largely suppressed in the large $N$ limit. For the analysis, we rule out an increase in the capacity of hormone secretions due to a larger $N$ by normalizing the hormone effectiveness ($\lambda N = 1$).

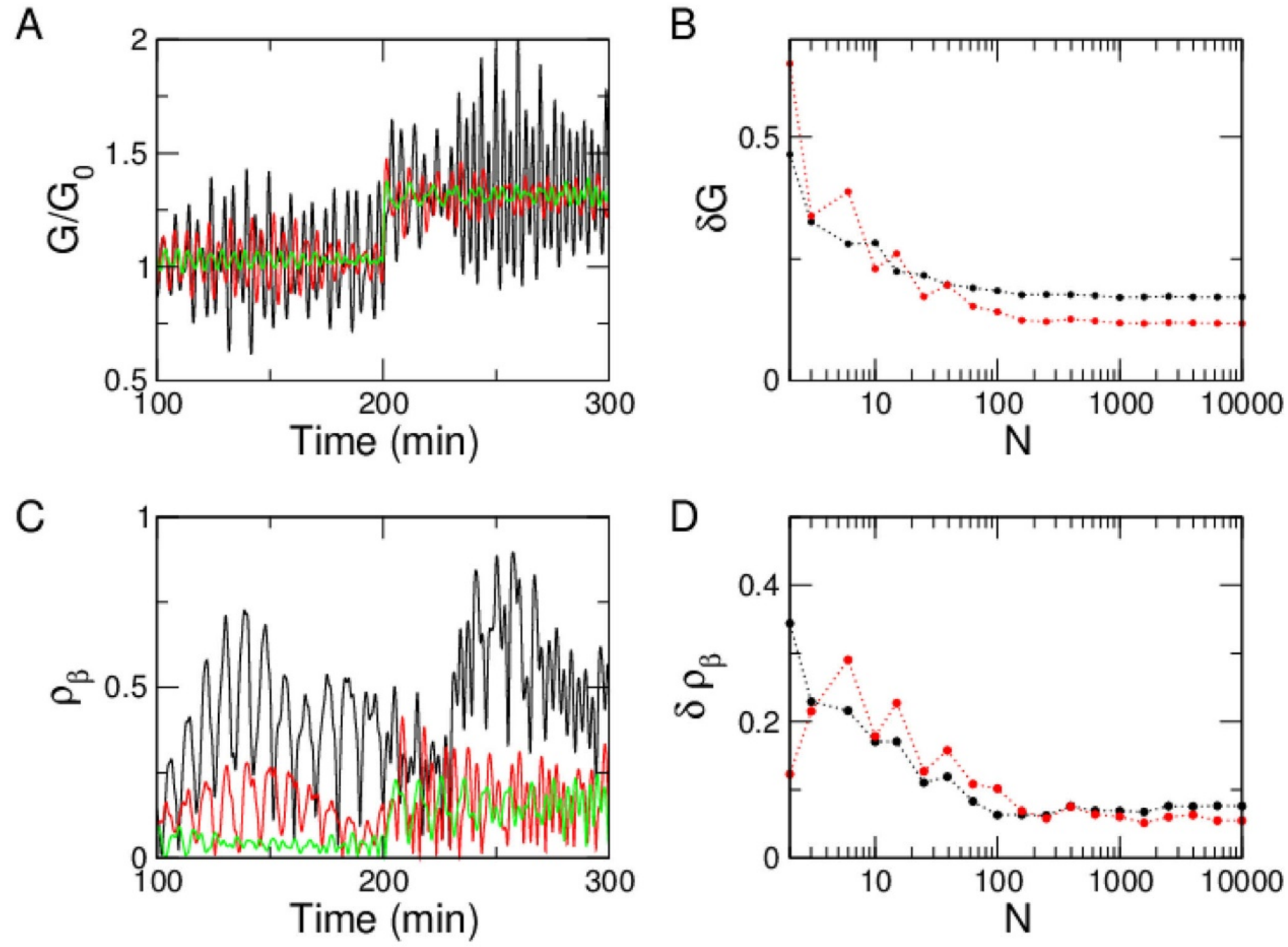

**Fig. S7. Effects of islet number.** (A) Temporal glucose profiles under a glucose stimulus ($I = 0.5G_0$) for Time > 200: $N$ =10 (black), 100 (red), and 1,000 (green). (B) Temporal fluctuations of glucose the concentration for various numbers of islets before (black) and after (red) the glucose stimulus. Given the same protocol, (C) inter-islet synchronization index for β cells, and (D) its fluctuations. Dotted lines are drawn for guiding eyes. For the simulation, network 121212 was used.

(6) Time delay of hormone actions, $\tau$

It takes time for the hormones secreted by the pancreas to act on the liver or peripheral tissues. The time delay can be simply considered by reformulating Eq. (2) as follows:

$$\dot{G} = \lambda N(G_0 H'_\alpha - G H'_\beta) + I(t), \qquad \text{(S9)}$$

$$\tau \dot{H}'_\sigma = H_\sigma - H'_\sigma, \qquad \text{(S10)}$$

where $\sigma \in \{\alpha, \beta\}$. The time delay, $\tau$, may have a time scale of 1 minute, considering the circulation time of blood in the body. Such a short delay does not affect the regulation of glucose (Fig. S8). However, a long delay ($\tau$ =10 min) could not effectively regulate glucose with less hormone secretions.

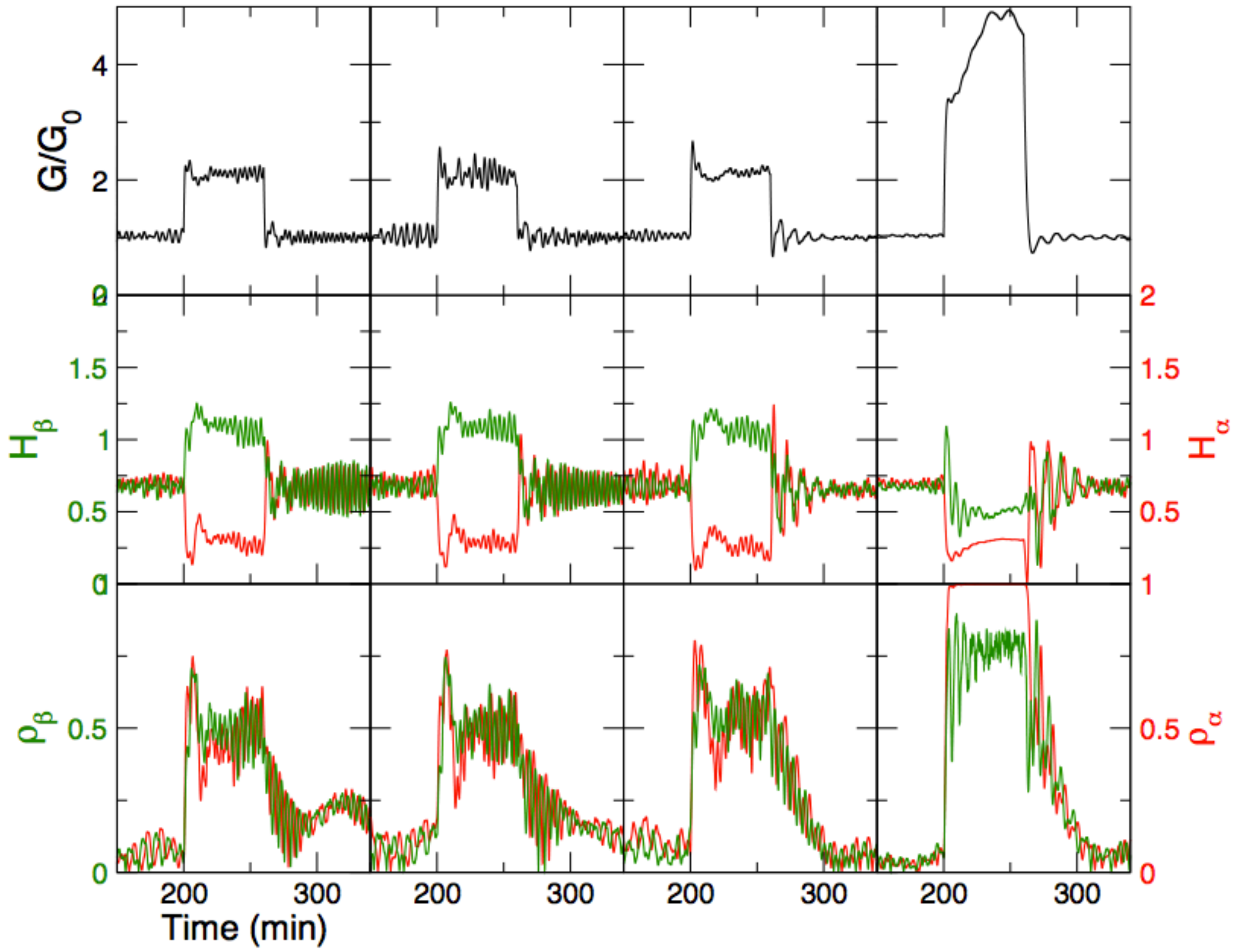

**Fig. S8. Time traces of glucose, hormones, and the degree of synchronization under time delays.** $\tau$ = 0, 0.1, 1 and 10 min from left to right. For the simulation, network 121212 was used with total $N$=200 islets.

### 3. Somatostatin oscillation and consumption

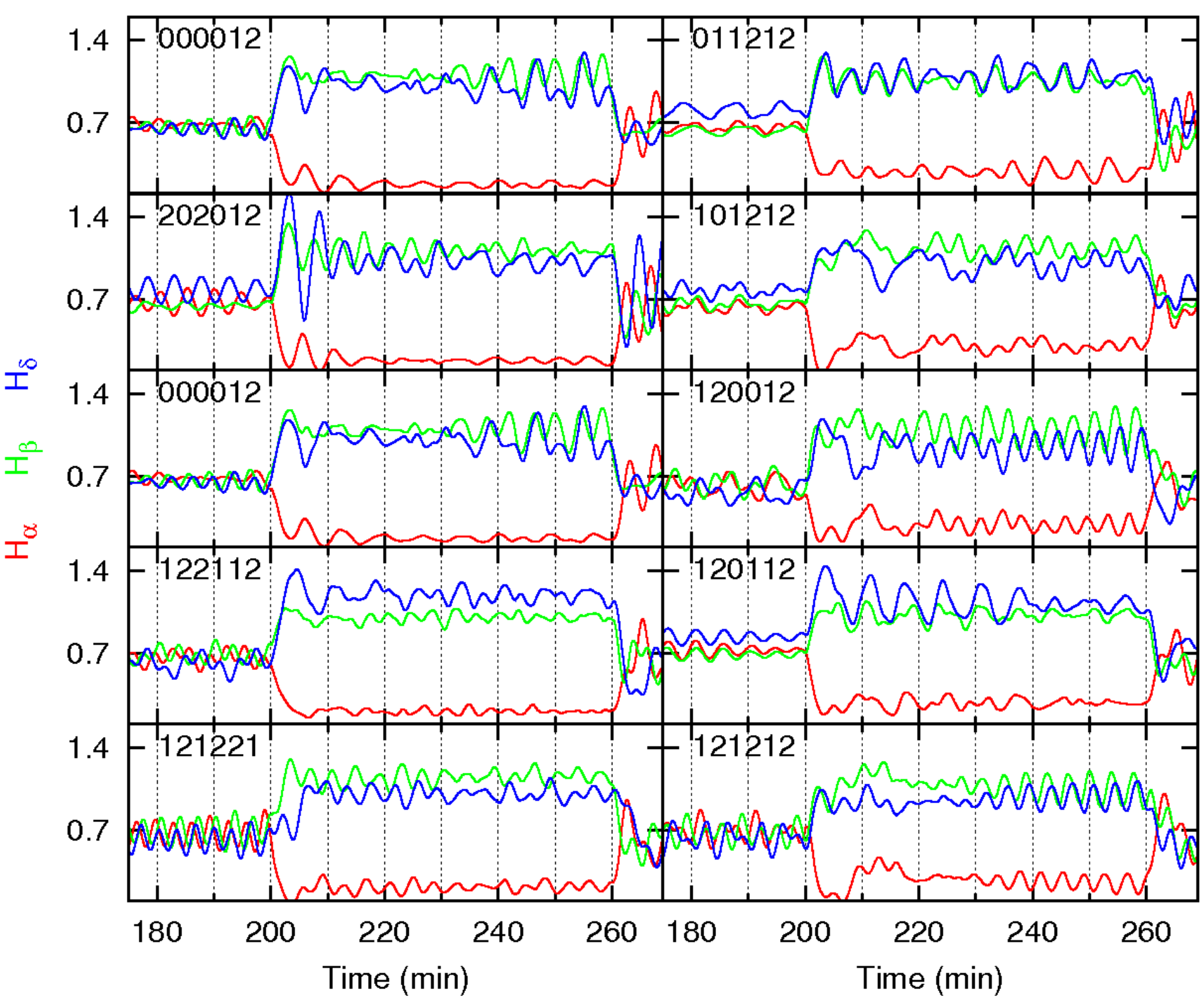

**Fig. S9. Temporal hormone profiles for the 10 effective networks.** Given the external glucose input $I = 2G_0$ for 200 < Time < 260, glucagon ($H_\alpha$, red), insulin ($H_\beta$, green), and somatostatin ($H_\delta$, blue) oscillated with time. Networks 121212 and 120012 showed clear phase coordination between the three hormones. For the simulation, standard parameter values were used (Table S1).

**Table S2. Total hormone consumption and controllability of inter-islet synchronization.**

| Rank | Network | $H_{\text{normal}}$ | $H_{\delta,\text{normal}}$ | $H_{\text{high}}$ | $H_{\delta,\text{high}}$ | $\Sigma H$ | $\rho_{\text{high}} - \rho_{\text{normal}}$ |
|---|---|---|---|---|---|---|---|
| 1 | **120012** | 1.249 | 0.575 | 1.388 | 0.942 | 4.154 | 0.307 |
| 2 | 101200 | 1.331 | 0.625 | 1.530 | 0.682 | 4.168 | 0.209 |
| 3 | 002102 | 1.312 | 0.626 | 1.574 | 0.711 | 4.223 | 0.209 |
| 4 | 202100 | 1.308 | 0.628 | 1.611 | 0.688 | 4.235 | 0.173 |
| 5 | 101202 | 1.423 | 0.621 | 1.528 | 0.677 | 4.249 | 0.206 |
| 6 | **121212** | 1.373 | 0.632 | 1.280 | 0.969 | 4.254 | 0.456 |
| 7 | **000012** | 1.347 | 0.651 | 1.293 | 0.996 | 4.287 | 0.228 |
| 8 | 210012 | 1.242 | 0.576 | 1.565 | 0.959 | 4.342 | 0.281 |
| 9 | **121221** | 1.363 | 0.612 | 1.376 | 0.994 | 4.345 | 0.323 |
| 10 | 002012 | 1.548 | 0.510 | 1.294 | 0.999 | 4.351 | 0.261 |
| 11 | 102102 | 1.512 | 0.611 | 1.533 | 0.717 | 4.373 | 0.238 |
| 12 | **122112** | 1.360 | 0.611 | 1.222 | 1.192 | 4.385 | 0.352 |
| 13 | 211202 | 1.529 | 0.602 | 1.468 | 0.798 | 4.397 | 0.290 |
| 14 | 102012 | 1.544 | 0.526 | 1.271 | 1.057 | 4.398 | 0.273 |
| 15 | 212100 | 1.258 | 0.594 | 1.558 | 1.011 | 4.421 | 0.316 |
| 16 | 120200 | 1.454 | 0.537 | 1.523 | 0.910 | 4.424 | 0.295 |
| 17 | **202012** | 1.333 | 0.774 | 1.298 | 1.026 | 4.431 | 0.214 |
| 18 | 122102 | 1.520 | 0.601 | 1.444 | 0.875 | 4.440 | 0.337 |
| 19 | 100000 | 1.616 | 0.548 | 1.435 | 0.867 | 4.466 | 0.215 |
| 20 | 002112 | 1.370 | 0.461 | 1.504 | 1.133 | 4.468 | 0.316 |
| 21 | 000002 | 1.561 | 0.559 | 1.488 | 0.864 | 4.472 | 0.252 |
| 22 | **100012** | 1.364 | 0.809 | 1.254 | 1.055 | 4.482 | 0.249 |
| 23 | **101212** | 1.284 | 0.772 | 1.421 | 1.013 | 4.490 | 0.276 |
| 24 | 120212 | 1.521 | 0.558 | 1.479 | 0.959 | 4.517 | 0.337 |
| 25 | 000112 | 1.307 | 0.626 | 1.559 | 1.052 | 4.544 | 0.285 |

Networks were sorted in ascending order of summed total hormone consumption at normal and high glucose conditions, $\Sigma H = H_{normal} + H_{\delta,normal} + H_{high} + H_{\delta,high}$, after selecting stable networks that showed small glucose fluctuations ($\delta\overline{G}^2_{\text{normal}} + \delta\overline{G}^2_{\text{high}} < 0.1^2$). The 10 effective networks in Table 1 have bold fonts.

## 4. Phase plane analysis for phase attractors

To understand the attractors of phase dynamics, we consider Eq. (S2) in a simplified setting in which the intrinsic angular velocities are identical ($\omega_{n\sigma} = \omega$), and the pinning term is off ($\mu = 0$). Note that the pinning term naturally becomes zero at normal glucose concentrations ($G = G_0$) in Eqs. (S6-8). Then, the phase equations for network 121212 are as follows:

$$\dot{\theta}_\alpha = \omega - K\frac{r_\beta}{r_\alpha}\sin(\theta_\beta - \theta_\alpha) - K\frac{r_\delta}{r_\alpha}\sin(\theta_\delta - \theta_\alpha), \qquad \text{(S12)}$$

$$\dot{\theta}_\beta = \omega + K\frac{r_\alpha}{r_\beta}\sin(\theta_\alpha - \theta_\beta) - K\frac{r_\delta}{r_\beta}\sin(\theta_\delta - \theta_\beta), \qquad \text{(S13)}$$

$$\dot{\theta}_\delta = \omega + K\frac{r_\alpha}{r_\delta}\sin(\theta_\alpha - \theta_\delta) + K\frac{r_\beta}{r_\delta}\sin(\theta_\beta - \theta_\delta). \qquad \text{(S14)}$$

Note that we removed the islet index, *n*, for simplicity. Because we are interested in the phase differences between α, β, and δ cells, relative phases ($x \equiv \theta_\alpha - \theta_\beta$ and $y \equiv \theta_\alpha - \theta_\delta$) can be defined. Using Eqs. (S12-14), we obtain the following:

$$\dot{x} = K\left[\frac{r_\beta}{r_\alpha} - \frac{r_\alpha}{r_\beta}\right]\sin x + K\frac{r_\beta}{r_\alpha}\sin y + K\frac{r_\delta}{r_\beta}\sin(x-y), \qquad \text{(S15)}$$

$$\dot{y} = K\frac{r_\beta}{r_\alpha}\sin x + \left[\frac{r_\delta}{r_\alpha} - \frac{r_\alpha}{r_\delta}\right]\sin y + K\frac{r_\beta}{r_\delta}\sin(x-y). \qquad \text{(S16)}$$

Depending on the amplitude, $r_\sigma$, different phase dynamics emerge, of approximately $r_\alpha > r_\beta, r_\delta$ at low glucose, $r_\beta > r_\alpha, r_\delta$ at high glucose, and $r_\alpha \approx r_\beta \approx r_\delta$ at normal glucose. Under the given glucose conditions, the phase dynamics exhibit single and triple attractors under low/high and normal glucose (Fig. S10).

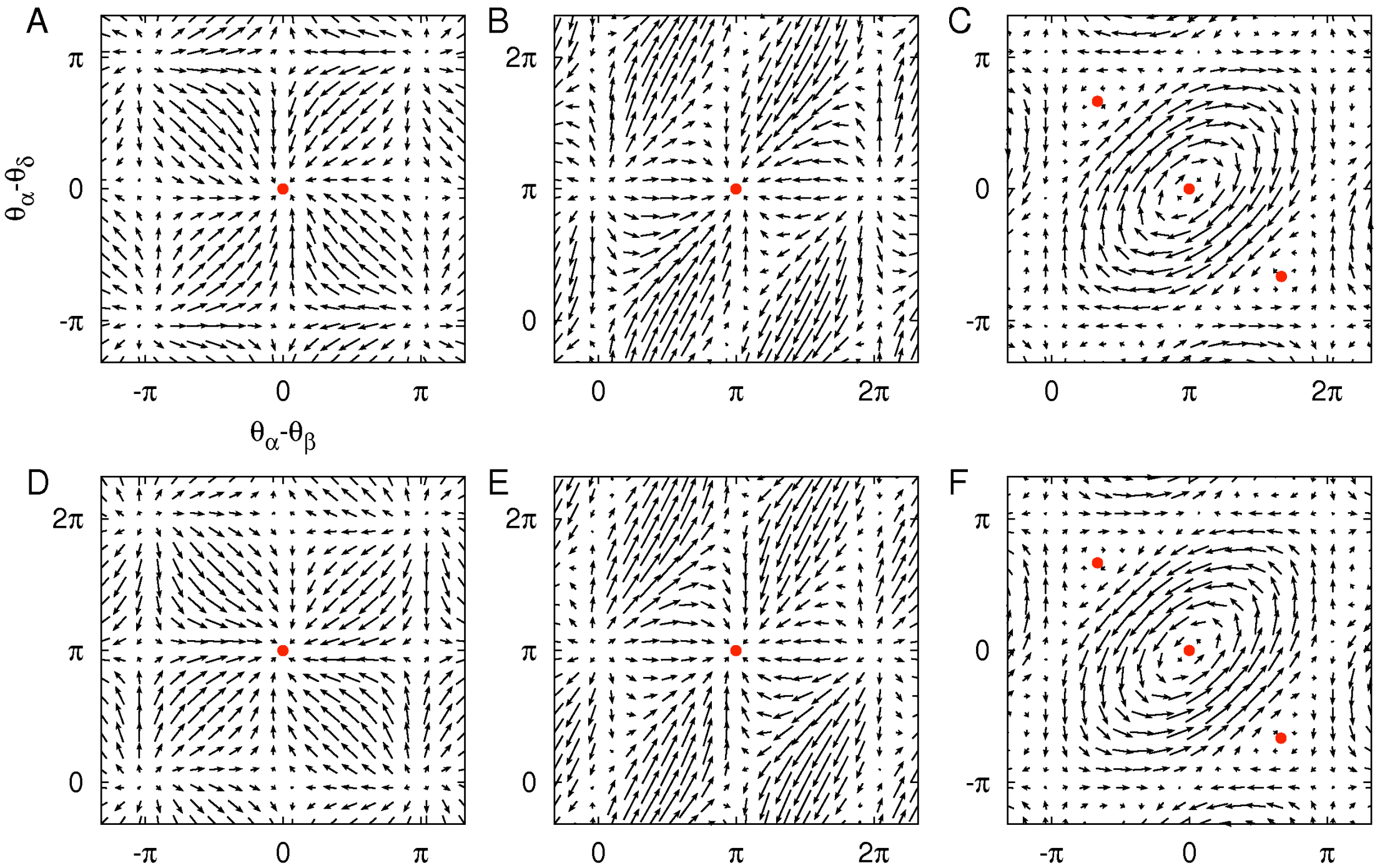

**Fig. S10. Vector flows in phase dynamics.** (A, D) $r_\alpha = 1, r_\beta = r_\delta = 0.2$, (B, E) $r_\beta = 1, r_\alpha = r_\delta = 0.2$, (C, F) $r_\alpha = r_\beta = r_\delta = 1$ for network 121212 (A, B, C) and network 122112 (D, E, F).

## 5. Low glucose challenge

Pancreatic islets regulate not only high glucose but also low glucose. Therefore, we challenged islets by lowering glucose concentrations with a negative glucose influx ( $I < 0$ ). Controllable inter-islet synchronization emerges at low glucose levels (Fig. S11).

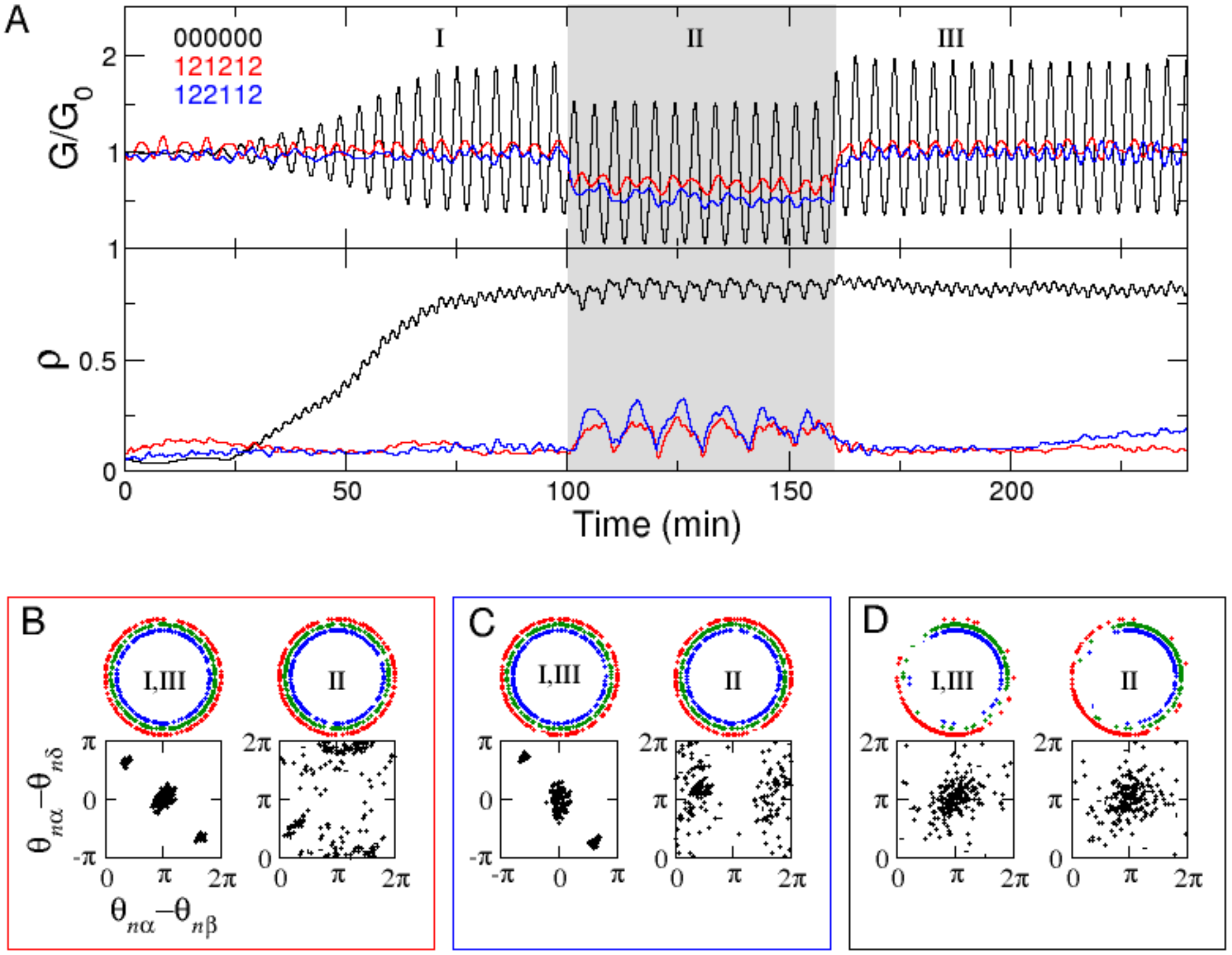

**Fig. S11. Controllable inter-islet synchronization and phase coordination between islet cells.** (A) Glucose regulation and inter-islet synchronization for networks 121212 (red), 122112 (blue) and 000000 (black), given the external glucose input ( $I = -0.5G_0$ ) during 100 < $t$ < 160. Phase snapshots of α, β, and δ cells under different glucose conditions (regimes I, II, and III) for (B) networks 121212, (C) 122112, and (D) 000000. Upper panel: absolute phases ( $\theta_{n\alpha}, \theta_{n\beta}, \theta_{n\delta}$ ) of α (red), β (green), and δ (blue) cells in the pancreas consisting of 200 islets. Lower panel: phase differences ( $\theta_{n\alpha} - \theta_{n\beta}, \theta_{n\alpha} - \theta_{n\delta}$ ). The axis range was adjusted to show distinct attractors considering $2\pi$ periodicity.

## 6. Population model

Real islets are composed of populations of islet cells rather than single α, β, and δ cells within each islet. Therefore, we consider populations of islet cells with a known composition and organization (*3*). In the population model, each cell has different nearest neighbors. The interaction follows exactly the same pattern as the single-cell model, but one has to consider autocrine interactions in which the same cell types interact with each other in this case. We used positive autocrine interactions, as in a previous study (*4*). We confirmed that controllable synchronization is also realized in the population model (Fig. S12).

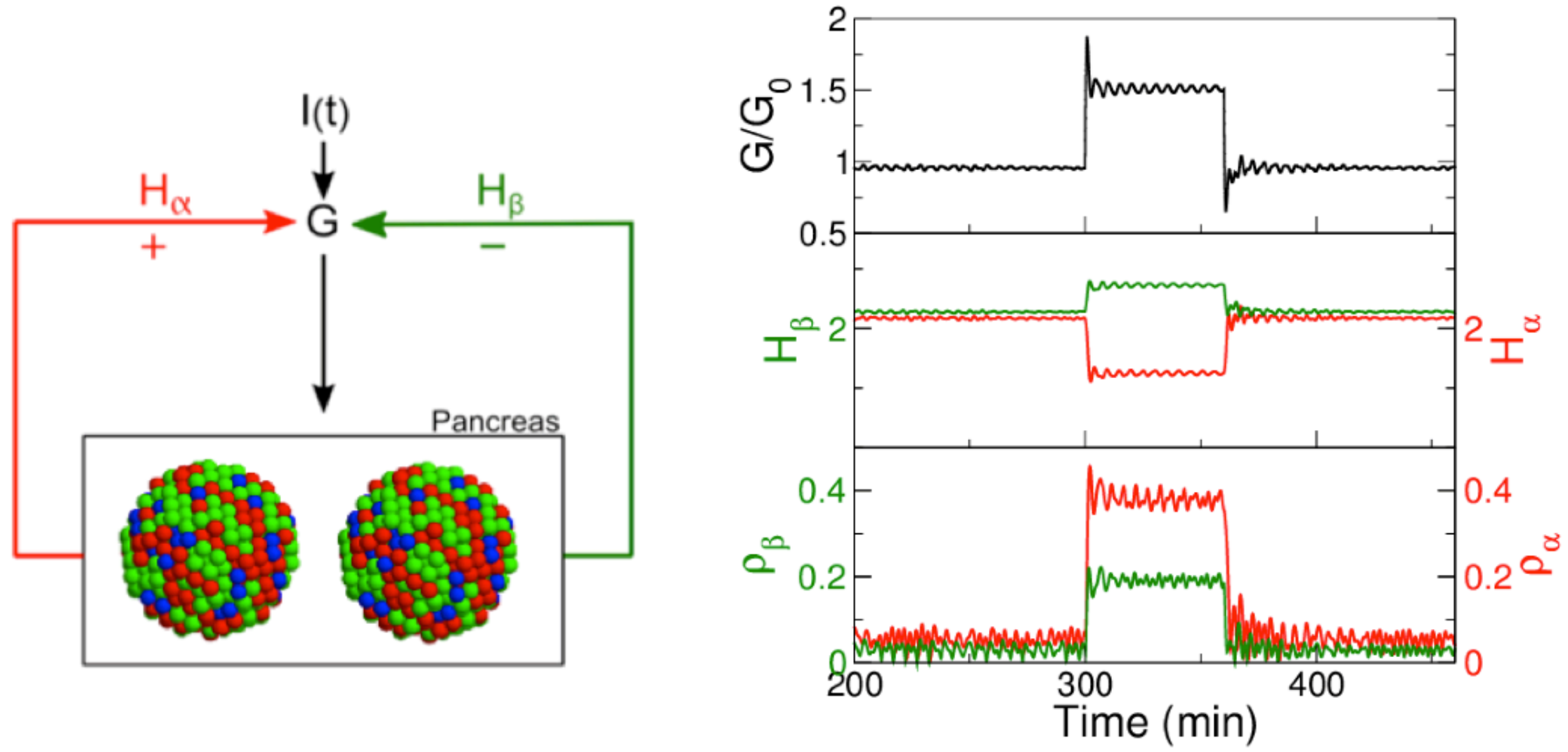

**Fig. S12. Population model and controllable synchronization.** The pancreas is composed of 1,000 islets, and each islet has 1,357 cells (30% α, 60% β, and 10% δ cells) that are organized for a partial mixing structure. Given glucose stimulus ($I = 3G_0$) for 300 < Time < 360, total 1,000×1,357 islet cells regulate glucose level. Plotted are glucose change, hormone secretions, and degrees of synchronization between α cells in different islets and between β cells in different islets.

### 7. Noisy glucose stimulus

To probe noise dependence for glucose regulation, we introduced noise in glucose input:

$$I(t) = I_1 + I_2 \xi(t), \qquad \text{(S17)}$$

where $\xi(t)$ is a white noise with $\langle \xi(t) \rangle = 0$ and $\langle \xi(t)\xi(t') \rangle = \delta(t-t')$. After conducting the stochastic simulation, we confirmed that the noise made glucose changes and synchronization indices more jiggling (Fig. S13), but it did not change our main conclusion about the effectiveness of network 121212 for glucose regulation (Fig. S14). However, we did not see exceptional noise tolerance of network 121212 compared with other network motifs.

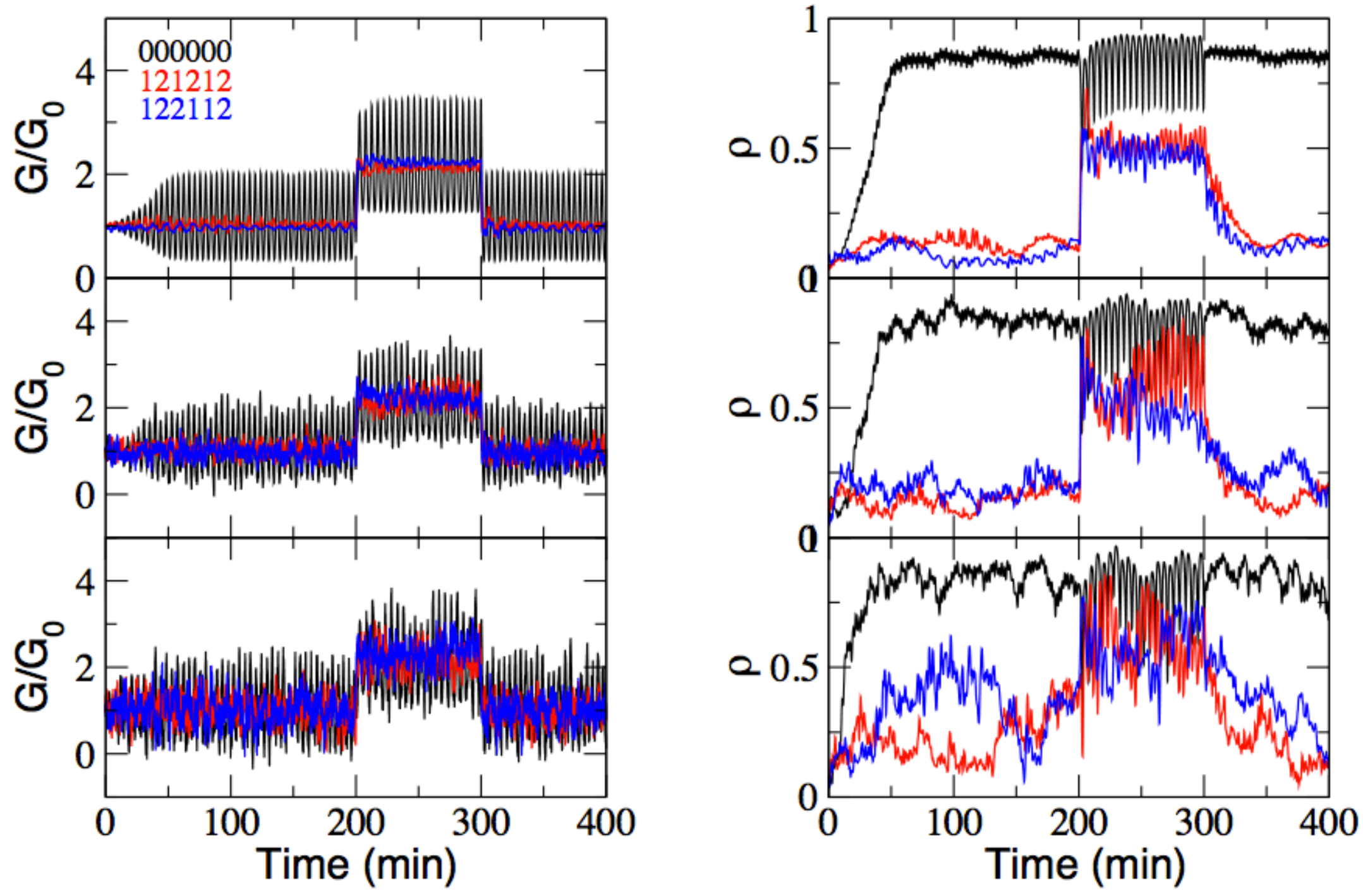

**Fig. S13. Noisy glucose stimulus.** Glucose (left) and inter-islet synchronization index (right) changes under noisy glucose stimuli: $I_2 = 0$ (top), $0.2G_0$ (middle), and $0.4G_0$ (bottom). To have normal and high glucose conditions, $I_1 = 2G_0$ was infused for $200 < \text{Time} < 300$, otherwise $I_1 = 0$. Networks 000000 (black), 121212 (red), and 122112 (blue). The standard parameter values were used for this simulation.

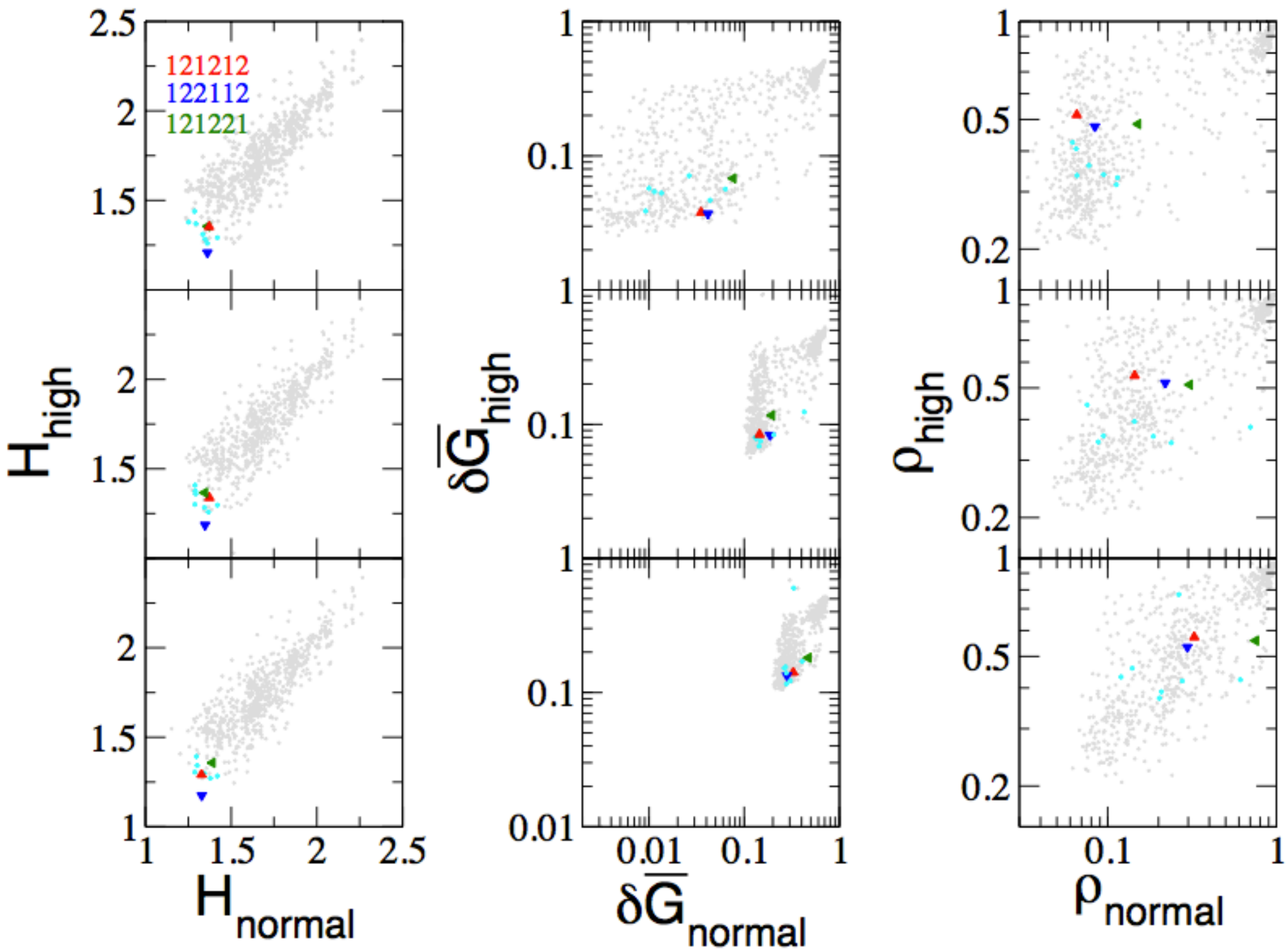

**Fig. S14. Effective networks under noisy conditions.** Hormone consumptions, glucose fluctuations, and inter-islet synchronizations under normal ($I_1 = 0$) and high ($I_1 = 2G_0$) glucose conditions with noisy glucose stimuli: $I_2 = 0$ (top), $0.2G_0$ (middle), and $0.4G_0$ (bottom). Networks 121212 (red), 122112 (blue), 121221 (green), other seven effective networks (cyan), and remaining 719 networks (gray). The standard parameter values were used for this simulation.

## 8. Oscillatory glucose stimulus

To probe the islet response under oscillatory glucose stimulus, we introduced an oscillating glucose input:

$$I(t) = I_1 \cos(\omega_{ext} t), \quad \text{(S17)}$$

where the external driving frequency was set as $2\pi / \omega_{ext} = 5$ min. If the driving frequency is too fast or too slow compared with the intrinsic frequencies $\omega_{n\sigma}$ of islet cells, the oscillatory stimuli cannot entrain islets. When the driving amplitude, $I_1$, was sufficiently large, the hormone secretions from islets were entrained to the external driving (Fig. S15). Here, networks 121212 and 122112 could resist to be entrained by small oscillatory glucose stimuli unlike network 000000.

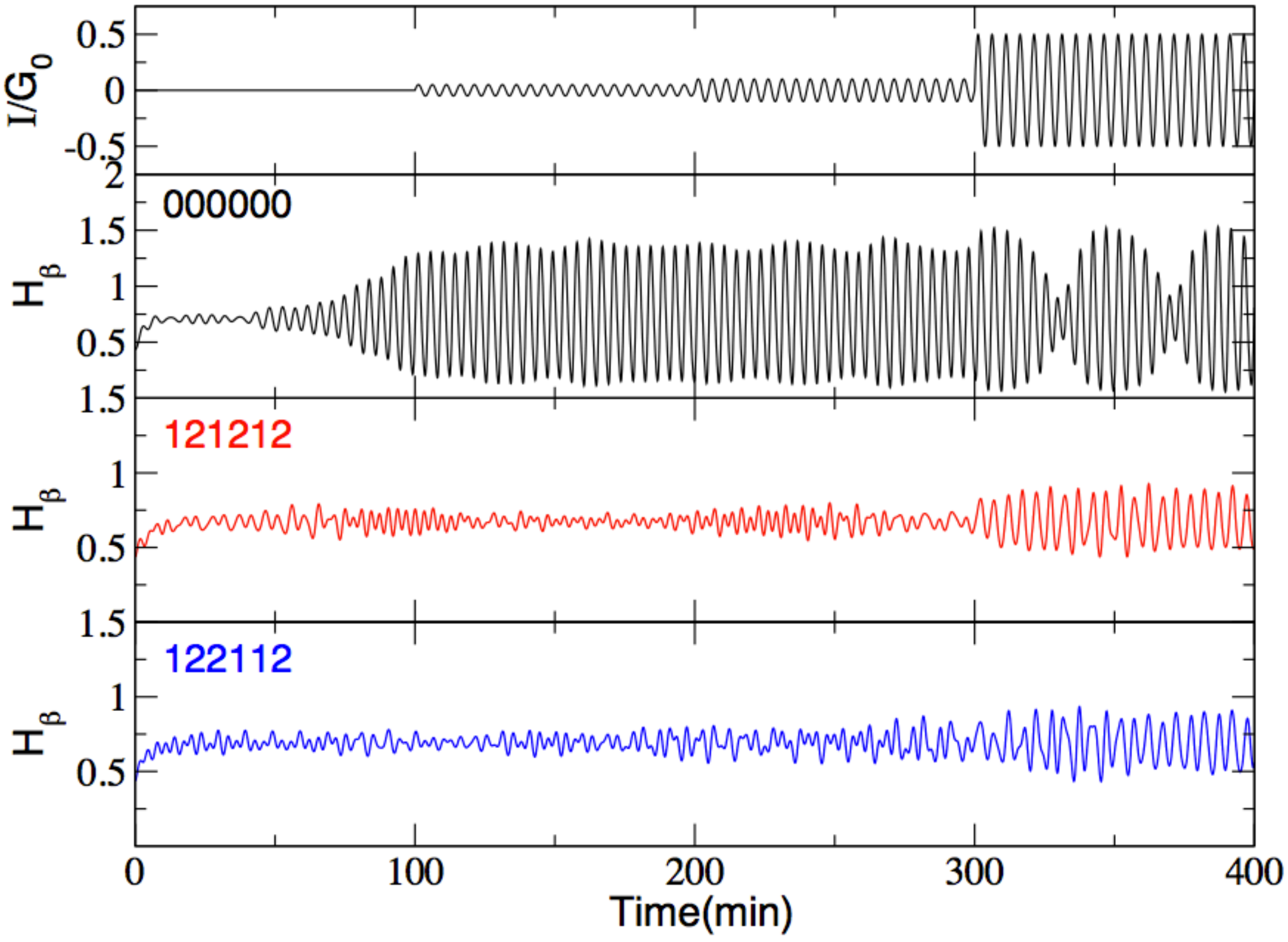

**Fig. S15. Oscillatory glucose stimulus and hormone secretion.** Hormone secretion of networks 000000 (black), 121212 (red), and 122112 (blue) under oscillatory glucose stimuli: $I_1 = 0$ (0 < Time < 100), $0.05G_0$ (100 < Time < 200), $0.1G_0$ (200 < Time < 300), and $0.5G_0$ (300 < Time < 400). The standard parameter values were used for this simulation.